\documentclass[longauth]{aa}
\usepackage[utf8]{inputenc}
\usepackage[varg]{txfonts}
\usepackage[usenames, dvipsnames]{color}
\usepackage{amsmath}
\usepackage{graphicx}
\usepackage[english]{babel}
\usepackage{siunitx}
\usepackage{float}
\usepackage{url}
\usepackage{longtable}
\usepackage{fancyhdr}
\usepackage{natbib}
\usepackage{booktabs}
\usepackage[normalem]{ulem}
\usepackage[version=4]{mhchem}
\usepackage{gensymb}
\usepackage[breaklinks, colorlinks, citecolor=blue, linkcolor=blue]{hyperref}
\makeatletter

\def\instrefs#1{{\def\scsep{\def\scsep{,}}\@for\w:=#1\do{\scsep\ref{inst:\w}}}}
\renewcommand{\inst}[1]{\unskip$^{\instrefs{#1}}$}
\def\instrefs#1{{\def\scsep{\def\scsep{,}}\@for\w:=#1\do{\scsep\ref{inst:\w}}}}
\renewcommand{\inst}[1]{\unskip$^{\instrefs{#1}}$}

\renewcommand{\autoref}
        {\def\equationautorefname{Eq.}%
         \def\figureautorefname{Fig.}%
         \def\sectionautorefname{Sect.}%
         \def\subsectionautorefname{Sect.}%
         \def\subsubsectionautorefname{Sect.}%
         \orgautoref}
\renewcommand*\aa@pageof{, page \thepage{} of \pageref*{LastPage}}
\makeatother

\usepackage[breaklinks, colorlinks, citecolor=blue, linkcolor=blue]{hyperref}
\usepackage{natbib}

\newcommand{\host}{G~9-40}
\newcommand{\planetb}{G~9-40~b}
\newcommand{\splx}{$\pi_{\star}$} 
\newcommand{\sdist}{$d_{\star}$}
\newcommand{\spmras}{$\mu_{\alpha}\cos\delta$}
\newcommand{\spmdec}{$\mu_{\delta}$}

\newcommand{\steff}{$T_\mathrm{eff}$}
\newcommand{\logsg}{$\log\,g_{\star}$}
\newcommand{\sfeh}{$\mathrm{[Fe/H]}$}

\newcommand{\sm}{$M_{\star}$}
\newcommand{\sr}{$R_{\star}$}

\newcommand{\slum}{$L_{\star}$}

\newcommand{\steffv}[1][$\mathrm{K}$]{${3395}\,{\pm}\,{51}$\,#1}
\newcommand{\logsgv}[1][$\mathrm{(cgs)}$]{${4.84}\,{\pm}\,{0.04}$\,#1}
\newcommand{\sfehv}[1][$\mathrm{dex}$]{${-0.07}\,{\pm}\,{0.16}$\,#1}
\newcommand{\smv}[1][$\mathrm{M_{\odot}}$]{${0.2952}\,{\pm}\,{0.0136}$\,#1}
\newcommand{\srv}[1][$\mathrm{R_{\odot}}$]{${0.3026}\,{\pm}\,{0.0095}$\,#1}
\newcommand{\slumv}[1][$10^{-4}$\,\Lsun]{${109.6}\,{\pm}\,{1.9}$\,#1}

\newcommand{\maspyr}{$\mathrm{mas\,yr^{-1}}$}
\newcommand{\Msun}{$\mathrm{M_{\odot}}$}
\newcommand{\Rsun}{$\mathrm{R_{\odot}}$}

\newcommand{\Lsun}{$\mathrm{L_{\odot}}$}
\newcommand{\mps}{$\mathrm{m\,s^{-1}}$}

\newcommand{\kmps}{$\mathrm{km\,s^{-1}}$}

\newcommand{\gaia}{\emph{\it Gaia}}

\newcommand{\jwst}{\emph{{\it JWST}}}

\begin{document} 
\title{Precise mass determination for the keystone sub-Neptune planet transiting the mid-type M~dwarf G~9-40}
\subtitle{}
\author{
R.~Luque\inst{uchicago,iaa}
\and G.~Nowak\inst{iac,ull}
\and T.~Hirano\inst{abc}
\and D.~Kossakowski\inst{mpia}
\and E.~Pall\'{e}\inst{iac,ull}
\and M.\,C.~Nixon\inst{ucambridge}
\and G.~Morello\inst{iac,ull}
\and P.\,J.~Amado\inst{iaa}
\and S.\,H.~Albrecht\inst{aarhus}
\and J.\,A.~Caballero\inst{cabesac}
\and C.~Cifuentes\inst{cabesac}
\and W.\,D.~Cochran\inst{mcdonald} 
\and H.\,J.~Deeg\inst{iac,ull}
\and S.~Dreizler\inst{iag}
\and E.~Esparza-Borges\inst{iac,ull}
\and A.~Fukui\inst{komaba,iac} 
\and D.~Gandolfi\inst{utorino}
\and E.~Goffo\inst{utorino,tls}
\and E.\,W.~Guenther\inst{tls}
\and A.\,P.~Hatzes\inst{tls}
\and T.~Henning\inst{mpia}
\and P.~Kabath\inst{ondrejov}
\and K.~Kawauchi\inst{iac,ull}
\and J.~Korth\inst{chalmers}
\and T.~Kotani\inst{naoj}
\and T.~Kudo\inst{subaru}
\and M.~Kuzuhara\inst{naoj}
\and M.~Lafarga\inst{warwick}
\and K.\,W.\,F.~Lam\inst{dlr}
\and J.~Livingston\inst{abc,naoj,sokendai}
\and J.\,C.~Morales\inst{ice,ieec}
\and A.~Muresan\inst{chalmers} 
\and F.~Murgas\inst{iac,ull}
\and N.~Narita\inst{komaba,abc,iac}
\and H.\,L.\,M.~Osborne\inst{ucl}
\and H.~Parviainen\inst{iac,ull}
\and V.\,M.~Passegger\inst{hs,oklahoma}
\and C.\,M.~Persson\inst{chalmers}
\and A.~Quirrenbach\inst{lsw}
\and S.~Redfield\inst{wesleyan}
\and S.~Reffert\inst{lsw}
\and A.~Reiners\inst{iag}
\and I.~Ribas\inst{ice,ieec}
\and L.\,M.~Serrano\inst{utorino}
\and M.~Tamura\inst{utokyo,abc,naoj} 
\and V.~Van~Eylen\inst{ucl}
\and N.~Watanabe\inst{dmds}
\and M.\,R.~Zapatero Osorio\inst{cab}
}

\institute{
\label{inst:uchicago}Department of Astronomy \& Astrophysics, University of Chicago, Chicago, IL 60637, USA; \email{rluque@uchicago.edu}
\and
\label{inst:iaa}Instituto de Astrof\'isica de Andaluc\'ia (IAA-CSIC), Glorieta de la Astronom\'ia s/n, 18008 Granada, Spain.
\and 
\label{inst:iac}Instituto de Astrof\'{i}sica de Canarias (IAC), 38205 La Laguna, Tenerife, Spain.
\and 
\label{inst:ull}Departamento de Astrof\'{i}sica, Universidad de La Laguna (ULL), 38206, La Laguna, Tenerife, Spain.
\and 
\label{inst:abc}Astrobiology Center, 2-21-1 Osawa, Mitaka, Tokyo 181-8588, Japan.
\and
\label{inst:mpia}Max-Planck-Institut für Astronomie, Königstuhl 17, 69117 Heidelberg, Germany.
\and
\label{inst:ucambridge}Institute of Astronomy, University of Cambridge, Madingley Road, Cambridge CB3 0HA, UK.
\and
\label{inst:aarhus}Stellar Astrophysics Centre, Department of Physics and Astronomy, Aarhus University, Ny Munkegade 120, 8000 Aarhus C, Denmark
\and
\label{inst:cabesac}Centro de Astrobiología (CSIC-INTA), ESAC, Camino bajo del castillo s/n, 28692 Villanueva de la Cañada, Madrid, Spain
\and
\label{inst:iag}Institut f\"ur Astrophysik, Georg-August-Universit\"at, Friedrich-Hund-Platz 1, 37077 G\"ottingen, Germany
\and
\label{inst:mcdonald}McDonald Observatory and Center for Planetary Systems Habitability, The University of Texas, Austin, Texas, USA.
\and 
\label{inst:komaba}Komaba Institute for Science, The University of Tokyo, 3-8-1 Komaba, Meguro, Tokyo 153-8902, Japan.
\and
\label{inst:utorino}Dipartimento di Fisica, Università degli Studi di Torino, via Pietro Giuria 1, I-10125, Torino, Italy.
\and
\label{inst:tls}Thüringer Landessternwarte Tautenburg, Sternwarte 5, 07778 Tautenburg, Germany.
\and
\label{inst:ondrejov}Astronomical Institute of the Czech Academy of Sciences, Fricova 298, 25165 Ondrejov, Czech Republic.
\and
\label{inst:chalmers}Department of Space, Earth and Environment, Astronomy and Plasma Physics, Chalmers University of Technology, 412 96 Gothenburg, Sweden.
\and 
\label{inst:naoj}National Astronomical Observatory of Japan, 2-21-1 Osawa, Mitaka, Tokyo 181-8588, Japan.
\and
\label{inst:subaru}Subaru Telescope, National Astronomical Observatory of Japan, 650 North A'ohoku Place, Hilo, HI 96720  USA. 
\and
\label{inst:warwick}Department of Physics, University of Warwick, Gibbet Hill Road, Coventry CV4 7AL, United Kingdom
\and
\label{inst:dlr}Institute of Planetary Research, German Aerospace Center (DLR), Rutherfordstrasse 2, 12489 Berlin, Germany.
\and 
\label{inst:sokendai}Department of Astronomy, The Graduate University for Advanced Studies (SOKENDAI), 2-21-1 Osawa, Mitaka, Tokyo, Japan.
\and 
\label{inst:ice}Institut de Ci\`{e}ncies de l'Espai (ICE, CSIC), Campus UAB, C/Can Magrans s/n, 08193 Bellaterra, Spain.
\and 
\label{inst:ieec}Institut d'Estudis Espacials de Catalunya (IEEC), C/ Gran Capit\`{a} 2-4, 08034 Barcelona, Spain.
\and 
\label{inst:ucl}Mullard Space Science Laboratory, University College London, Holmbury St Mary, Dorking, Surrey RH5 6NT, UK
\and
\label{inst:hs}Hamburger Sternwarte, Gojenbergsweg 112, D-21029 Hamburg, Germany.
\and
\label{inst:oklahoma}Homer L. Dodge Department of Physics and Astronomy, University of Oklahoma, 440 West Brooks Street, Norman, OK 73019, USA.
\and 
\label{inst:lsw}Landessternwarte, Zentrum für Astronomie der Universität Heidelberg, Königstuhl 12, 69117 Heidelberg, Germany.
\and
\label{inst:wesleyan}Astronomy Department and Van Vleck Observatory, Wesleyan University, Middletown, CT 06459, USA.
\and
\label{inst:iag}Institut für Astrophysik und Geophysik, Georg-August-Universität Göttingen, Friedrich-Hund-Platz 1, 37077 Göttingen, Germany
\and 
\label{inst:utokyo}Department of Astronomy, University of Tokyo, 7-3-1 Hongo, Bunkyo-ku, Tokyo 113-0033, Japan.
\and
\label{inst:dmds}Department of Multi-Disciplinary Sciences, Graduate School of Arts and Sciences, The University of Tokyo, 3-8-1 Komaba, Meguro, Tokyo 153-8902, Japan.
\and
\label{inst:cab}Centro de Astrobiología (CSIC-INTA), Carretera de Ajalvir km 4, 28850 Torrejón de Ardoz, Madrid, Spain
}

\date{Received 05 July 2022 / Accepted 12 August 2022}

\abstract{Despite being a prominent subset of the exoplanet population discovered in the past three decades, the nature and provenance of sub-Neptune-sized planets is still one of the open questions in exoplanet science.} 
{For planets orbiting bright stars, precisely measuring the orbital and planet parameters of the system is the best approach to distinguish between competing theories regarding their formation and evolution.} 
{We obtained 69 new radial velocity observations of the mid-M~dwarf G~9-40 with the CARMENES instrument to measure for the first time the mass of its transiting sub-Neptune planet, G~9-40~b, discovered in data from the \textit{K2} mission.} 
{Combined with new observations from the \textit{TESS} mission during Sectors 44, 45, and 46, we are able to measure the radius of the planet to an uncertainty of 3.4\% ($R_{\rm b} = 1.900\pm0.065\,R_\oplus$) and determine its mass with a precision of 16\% ($M_{\rm b} = 4.00\pm0.63\,M_\oplus$). The resulting bulk density of the planet is inconsistent with a terrestrial composition and suggests the presence of either a water-rich core or a significant hydrogen-rich envelope. } 
{G~9-40~b is referred to as a keystone planet due to its location in period-radius space within the radius valley. Several theories offer explanations for the origin and properties of this population and this planet is a valuable target for testing the dependence of those models on stellar host mass. By virtue of its brightness and small size of the host, it joins L~98-59~d as one of the two best warm ($T_{\rm eq} \sim 400\,\mathrm{K}$) sub-Neptunes for atmospheric characterization with \textit{JWST}, which will probe cloud formation in sub-Neptune-sized planets and break the degeneracies of internal composition models. } 


\maketitle

\section{Introduction}

The mass of an exoplanet is its most fundamental property. When it is combined with the radius of planets detected to be transiting, one can determine the planet's bulk density and constrain its bulk composition to a first approximation. Moreover, the mass of an exoplanet is a key quantity that also gives an insight on its formation and evolution history. For example, the exoplanet mass function can constrain the initial conditions of planet formation models and discriminate between different evolution scenarios \citep[e.g.,][]{Suzuki2018}. Precise mass measurements are crucial in order to correctly interpret exoplanetary atmospheric spectroscopic observations in emission or transmission. Uncertainties better than 20\,\% for the planet mass are necessary to retrieve accurate atmospheric properties \citep{Batalha2019}, especially for planets smaller than Neptune ($R < 4\,R_\oplus$), typically referred to as small planets. 

The \textit{Kepler} mission \citep{Kepler} demonstrated that planets with sizes between that of Earth and Neptune, with no counterpart in our Solar System, orbit about half of all planetary systems with Sun-like hosts \citep{Howard2012,Fressin2013}. The origin and nature of these planets are still under debate \citep[see the review by][]{Bean2021}, but the bimodality of their size distribution --- with a dearth of planets with radii between 1.7 and 2.0\,$R_\oplus$ \citep{Fulton2018,VanEylen2018} for solar-type stars and between 1.5 and 1.7\,$R_\oplus$ \citep{CloutierMenou2020,HardegreeUllman2020,VanEylen2021} for M-dwarf stars --- indicates that their composition ranges between scaled-up versions of Earth and the scaled-down versions of Neptune. These planets could be made of rock, ices, and/or gas, since multiple combinations of these materials can match the observed mean densities \citep[e.g.,][]{RogersSeager2010ApJ...712..974R}. Only by combining precise bulk density measurements with elemental stellar abundances is it possible to break the degeneracies of the internal composition models to derive reliable rock, water, and atmospheric mass fractions \citep[e.g.,][]{Delrez2021,Caballero2022arXiv220609990C}.

Atmospheric mass loss mechanisms such as photoevaporation \citep{OwenWu2017ApJ...847...29O,JinMordasini2018ApJ...853..163J} or core-powered mass loss \citep{Ginzburg2018MNRAS.476..759G,GuptaSchlichting2019MNRAS.487...24G} are both able to adequately reproduce the bimodal size distribution of small planets assuming that their cores are rocky in composition. This has led to the interpretation that super-Earths and sub-Neptunes have formed accreting only dry condensates within the water ice line, and their difference in sizes is attributed to the absence or presence of extended volatile-rich atmospheres. However, given their bulk densities, another possibility is that sub-Neptune planets are water-rich worlds. Their existence has been hypothesized by many authors \citep[e.g.,][]{Kuchner2003ApJ...596L.105K,Leger2004Icar..169..499L,Selsis2007Icar..191..453S,Sotin2007Icar..191..337S}
and they are a natural outcome of self-consistent planet formation and evolution models \citep[e.g.,][]{Raymond2018,Bitsch2019,Venturini2020,Bitsch2021,Izidoro2021}. Additionally, water worlds are able to explain observational features in the mass-radius diagram \citep{Zeng2019PNAS..116.9723Z,Mousis2020,Venturini2020} and the atmospheric composition of sub-Neptunes derived from recent transmission spectroscopy observations \citep{Benneke2019a,Benneke2019b,Tsiaras2019,Kreidberg2020,GarciaMunoz2021}. Only by enlarging the sample of precisely characterized planets amenable for follow-up atmospheric observations we will be able to distinguish between these two scenarios.

Here, we present the characterization of a sub-Neptune planet on a 5.75-day orbit around the relatively bright ($V=13.8$\,mag, $J=10.1$\,mag) M2.5 dwarf \host{}. The planet was first identified as a planet candidate by \citet{Yu2018AJ....156...22Y} and statistically validated as a planet by \citet{Stefansson2020AJ....159..100S}. Our radial velocity (RV) follow-up of the star using CARMENES \citep{CARMENES2020} enables us now to obtain a mass measurement, and thus constrain the bulk density of the planet with an uncertainty better than 20\,\%, pointing towards a water world composition or the presence of a significant volatile-rich atmosphere. Its location in the period-radius diagram where atmospheric mass-loss and gas-poor formation models make opposite predictions \citep[see e.g.,][]{CloutierMenou2020,Luque2021,Cloutier2021} makes it an interesting target to test theories about the nature of the radius valley.

\section{Observations}
\label{sec-photometry}

\subsection{Space-based photometry}
\label{subsec-photometry-space-based}

\begin{figure*}[ht!]
    \centering
    \includegraphics[width=0.96\hsize]{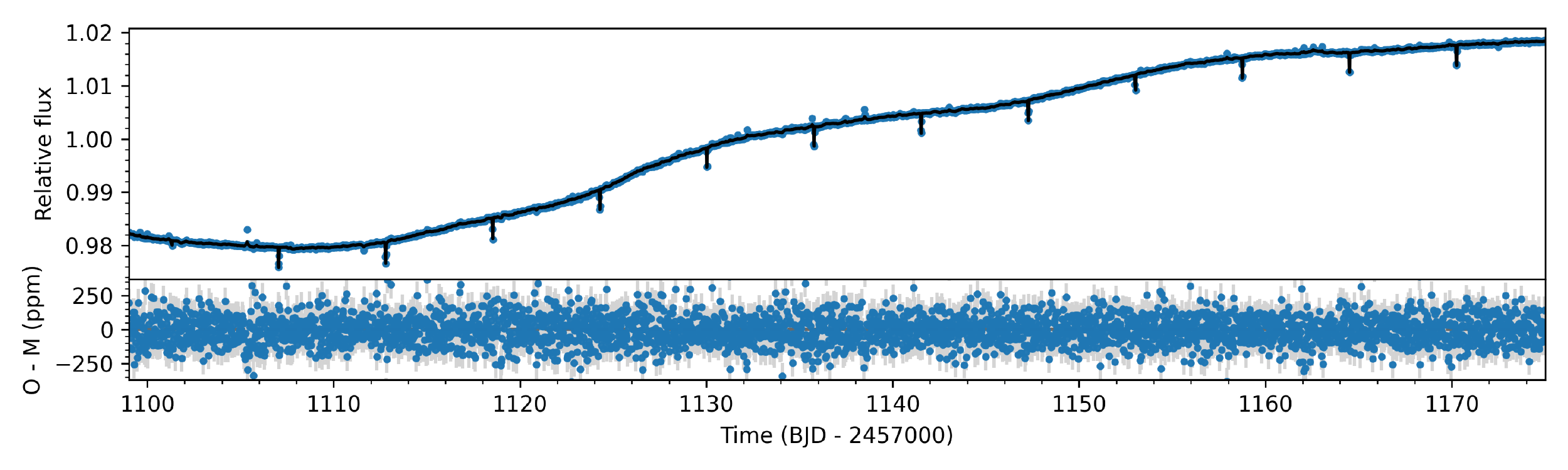}\\[0.1cm]
    \includegraphics[width=0.96\hsize]{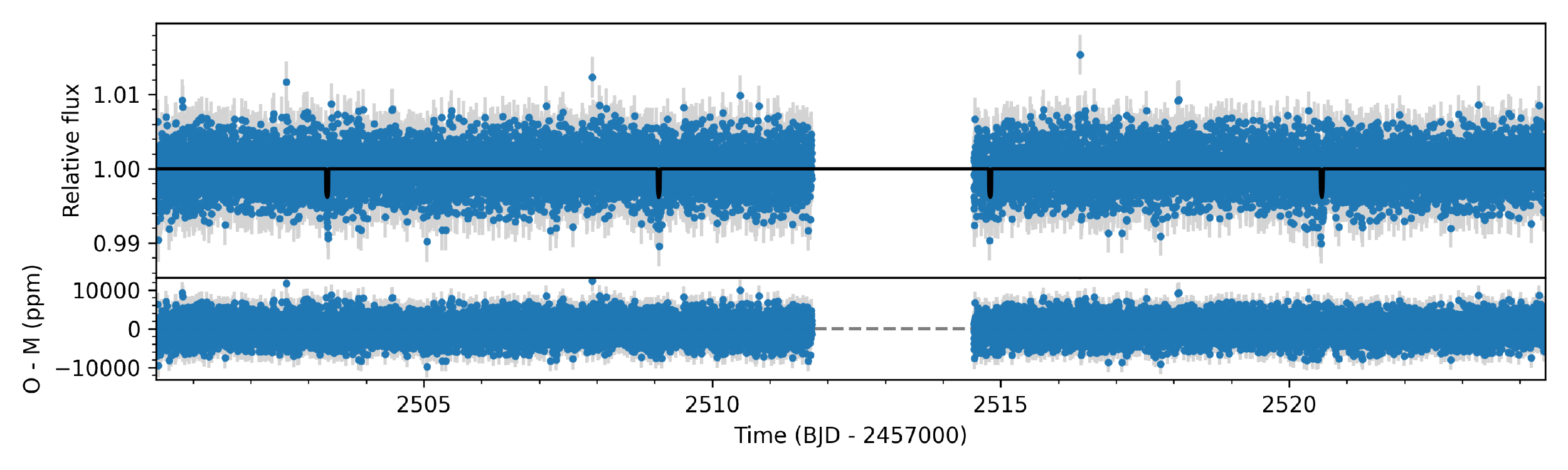}\\
    \includegraphics[width=0.96\hsize]{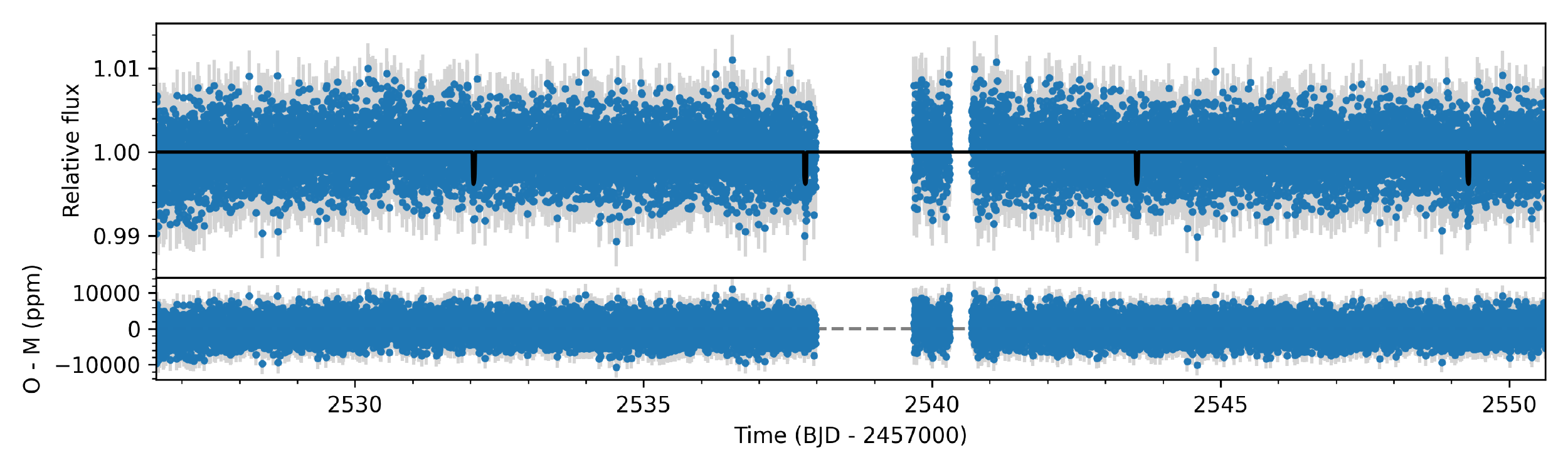}\\
    \includegraphics[width=0.96\hsize]{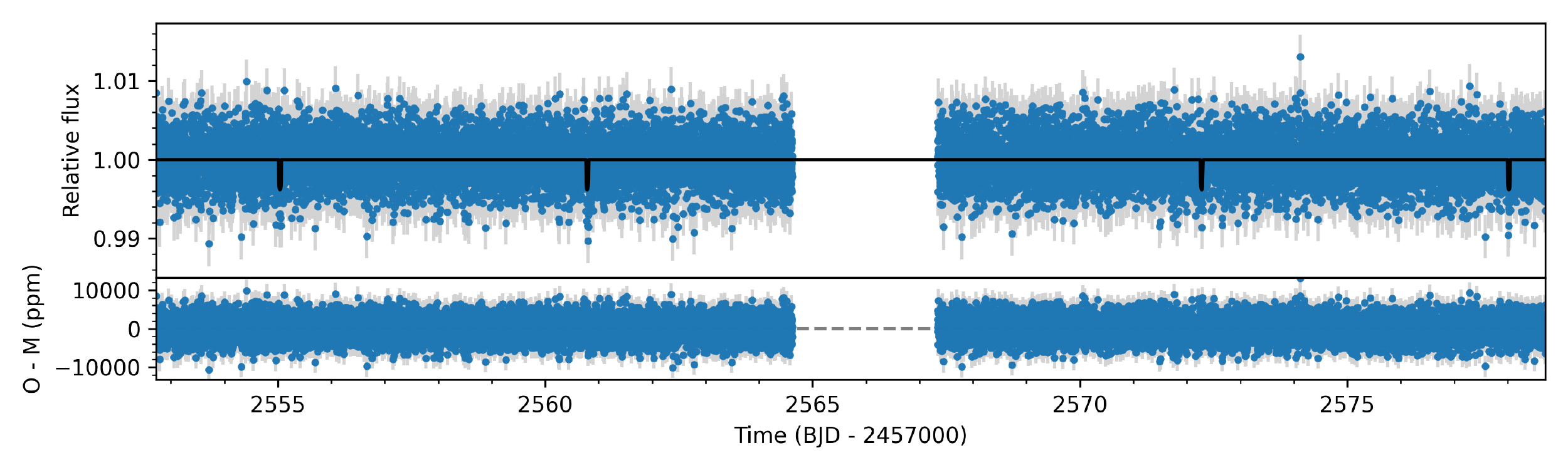}
    \caption{Space-based photometry of \host{}. \textit{K2} data are shown in the first panel, \textit{TESS} PDC-corrected SAP transit photometry from Sectors 44, 45, and 46 is shown in the second, third, and forth rows, respectively. The black line shows the joint fit to the data using \texttt{juliet}, including stellar variability (see Sect.~\ref{subsec-analysis-joint_model} for details on the modeling).}
    \label{fig:tess_lc}
\end{figure*}

\subsubsection{\emph{K2}}
\label{subsubsec-photometry-space-based-k2}
\host{} (EPIC 212048748) was observed by the \textit{K2} mission during Campaign 16 as part of the guest observer programs GO16005 (PI: Crossfield), GO16009 (PI: Charbonneau), GO16052 (PI: Stello), and GO16083 (PI: Coughlin). The star was monitored at 30-minute cadence from 7 December 2017 to 25 February 2018. 
To build the light curve from the target pixel file\footnote{Target pixel file was downloaded from the Mikulski Archive for Space Telescopes (MAST; \url{https://mast.stsci.edu})}, we used a method based on the implementation of the pixel level decorrelation (PLD) model \citep{Deming2015ApJ...805..132D} using a modified and updated version of the \texttt{Everest} pipeline \citep{Luger2018AJ....156...99L}. The details of this procedure are described in \citet{Palle2019A&A...623A..41P} and \citet{Hidalgo2020A&A...636A..89H}. In short, it extracts the raw light curve using a customized aperture that includes all the pixels around the photocenter of the star with signal above 1.7$\sigma$ of a previously calculated background. To perform robust flat-fielding corrections it removes all time cadences flagged as bad-quality data and then applies the PLD to the data up to third order, which does not require solving for correlations on stellar position. Finally, to separate astrophysical and instrumental variability it uses a second step of Gaussian Processes (GP). The resulting, detrended light curve preserving stellar variability is shown in Fig.~\ref{fig:tess_lc}. 

\subsubsection{TESS}
\label{subsubsec-photometry-space-based-tess}
\host{} (TIC-203214081) was observed by TESS at 2-minute cadence from 12 October 2021, until 30 December 2021, in sectors~44, 45, and 46, using cameras \#4, \#3, and \#1, respectively. The star was included in the Cycle 4 Guest Investigator programs G04039 (PI: Davenport), G04098 (PI: Kane), G04148 (PI: Robertson), G04205 (PI: Rodriguez), and G04242 (PI: Mayo), and it will not be observed again during the extended mission. The Science Processing Operations Center \citep[SPOC;][]{SPOC} at the NASA Ames Research Center made the data available at the MAST. SPOC provided simple aperture photometry (SAP) for this target as well as systematics-corrected photometry, a procedure consisting of an adaptation of the Kepler Presearch Data Conditioning algorithm \citep[PDC;][]{Smith2012PASP..124.1000S, Stumpe2012PASP..124..985S, Stumpe2014PASP..126..100S} to TESS. For the remainder of this work, we make use of the PDC-corrected SAP data, as shown in Fig.~\ref{fig:tess_lc}.

\subsection{Ground-based photometry}
\label{subsec-photometry-ground-based}

\subsubsection{MuSCAT2}
We observed one partial transit of \planetb{} on 14 May 2019, with the MuSCAT2 multi-color imager \citep{MuSCAT2} installed at the  Telescopio Carlos S\'anchez (TCS) located at Teide Observatory in Tenerife, Spain. Observations were carried out simultaneously in three bands ($r$, $i$, $z_s$), with a pixel scale of 0\farcs44\,pix$^{-1}$. We set the exposure times to avoid saturation of the target star and reduced the data using standard procedures. The photometry and transit model fit (including a linear baseline model with the airmass, seeing, x- and y-centroid shifts, and the sky level as covariates) was performed using the MuSCAT2 pipeline \citep{Parviainen2019A&A...630A..89P,Parviainen2020A&A...633A..28P}.

\subsubsection{ARCTIC}
We added to our analysis the full transit of \planetb{} observed using the ARCTIC imager \citep{ARCTIC} on the ARC 3.5-m Telescope at Apache Point Observatory on 14 April 2019. Details about the data reduction and analysis are presented in \citet{Stefansson2020AJ....159..100S}.

\subsection{High-contrast imaging}
\label{subsec-observations-hci}

\begin{figure}
\centerline{\includegraphics[angle=0,width=\linewidth]{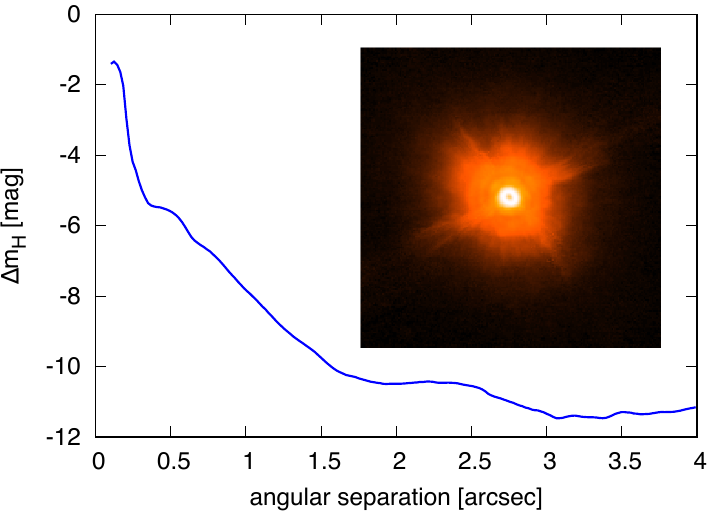}}
\caption{$5\sigma$ contrast curve against angular separation from \host, based on the Subaru/IRCS AO imaging. The inset displays a $4^{\prime\prime} \times 4^{\prime\prime}$ image around the star.
\label{figure-5sigma_contrast_epic212048748_inset}}
\end{figure}

To search for faint nearby stars and estimate a potential contamination factor from such sources we used a high contrast image of \host{} acquired on 29 March 2018, using the Subaru 8.2-m telescope and its adaptive optics (AO) system facility with the InfraRed Camera and Spectrograph \citep[IRCS,][]{IRCS}. Adopting the target itself as a natural guide for AO, we imaged it in the $H$ band with a five-point dithering. We obtained both short-exposure (unsaturated; $0.5\,\mathrm{s} \times 3$ co-addition for each dithering position) and long-exposure (mildly saturated; $2.0\,\mathrm{s} \times 3$ coaddition for each) frames of the target for absolute flux calibration and for inspecting nearby faint sources, respectively. We reduced the IRCS data following \citet{2016ApJ...820...41H} and obtained the median-combined images for unsaturated and saturated frames, respectively. Based on the unsaturated image, we estimated the target's full width at half maximum (FWHM) to be $0\farcs115$. Visual inspection of the saturated image suggests no nearby companion within $5^{\prime\prime}$ from \host. In order to estimate the detection limit of nearby faint companions around \host, we computed the $5\sigma$ contrast as a function of angular separation based on the flux scatter in each small annulus from the saturated target. Our AO imaging achieved approximately $\Delta H^\prime=8$\,mag at $1^{\prime\prime}$ from the star. Figure \ref{figure-5sigma_contrast_epic212048748_inset} plots the $5\sigma$ contrast along with the $4^{\prime\prime} \times 4^{\prime\prime}$ image in the inset.

\subsection{High-resolution spectroscopy}
\label{sec:obs-hrs}

High-resolution spectroscopic observations of \host{} were obtained between 8 November 2018, and 19 April 2021, using the CAHA3.5m/CARMENES and Subaru/IRD spectrographs. We collected a total of 69 CARMENES and 13 IRD spectra, which we combined with the existing 8 HET/HPF spectra from \citet{Stefansson2020AJ....159..100S}. 

\subsubsection{CARMENES}
We observed \host{} using the CARMENES instrument installed at the 3.5-m telescope Calar Alto Observatory, Almer\'ia, Spain, under the observing programs F19-3.5-014 (PI: Nowak) and F20-3.5-011 (PI: Nowak). The CARMENES spectrograph has two arms \citep{CARMENES2020}, the visible (VIS) arm covering the spectral range 0.52--\SI{0.96}{\micro\metre} and a near-infrared (near-IR) arm covering the spectral range 0.96--\SI{1.71}{\micro\metre}. Here, we use only the VIS channel observations to derive RV measurements since the intrinsic median error of the NIR RVs ($> 12$\,\mps{}) is much larger than the upper limit on the planet's semi-amplitude ($< 9$\,\mps{}) measured by \citet{Stefansson2020AJ....159..100S}. All observations were taken with exposure times of 1800\,s resulting in a signal-to-noise ratio (S/N) per spectral sample at 745\,nm in the range 25--49. The performance of the CARMENES instrument, the data reduction, and the wavelength calibration are described in \citet{2018A&A...609A.117T} and \citet{2018A&A...618A.115K}. Relative RV values, chromatic index (CRX), differential line width (dLW), and spectral index values such as $\alpha$ and CaII\,IRT were obtained using {\tt serval}\footnote{\url{https://github.com/mzechmeister/serval}} \citep{2018A&A...609A..12Z}. For each spectrum, we also computed the cross-correlation function (CCF) and its full-width at half maximum (FWHM), contrast (CTR) and bisector velocity span (BIS) values, using \texttt{raccoon}\footnote{\url{https://github.com/mlafarga/raccoon}} \citep{2020A&A...636A..36L}. The RV measurements were corrected for barycentric motion, secular acceleration and nightly zero-points \citep{Trifonov2018A&A...609A.117T}. The mean internal uncertainty is 2.3\,\mps{} for the RVs derived with \texttt{serval} and 4.7\,\mps{} for the RVs derived with \texttt{raccoon}. We use the \texttt{serval}-derived measurements in the following analyses.

\subsubsection{IRD}
We observed \host{} with the InfraRed Doppler instrument \citep[IRD,][]{Tamura2012SPIE.8446E..1TT,2018SPIE10702E..11K} behind an AO system \citep[AO188,][]{2010SPIE.7736E..0NH} on the Subaru 8.2-m telescope on Mauna Kea Observatories, as part of the open-use programs for following up transiting planet 
candidates (S18B-114 and S19A–069I). 
We took a total of 13 spectra between January 2019 and April 2021, simultaneously with laser-frequency comb spectra.
Exposure times were typically 600-1200\,s and the S/N at $\SI{1.0}{\micro\metre}$ was between 60--80 per spectral bin. We reduced the raw IRD frames of \host{} using the Échelle package of {\tt iraf} for flat-fielding, scattered-light subtraction, aperture tracing, and wavelength calibration with the Th-Ar lamp spectra. For a more precise wavelength calibration, the wavelength was re-calibrated based on the emission lines of the combined laser frequency comb, which was injected simultaneously into both stellar and reference fibres during instrument calibrations. We injected these reduced spectra into the RV analysis pipeline for Subaru/IRD \citep{2020ApJ...890L..27H} and attempted to reproduce the intrinsic stellar template spectrum from all the observed spectra with instrumental profile deconvolution and telluric removal. RVs were measured with respect to that template by forward-modeling of the observed individual spectral segments (each spanning 0.7--1.0\,nm). The uncertainties of the measured RVs are in the range 3.1--6.8\,\mps{} with a mean value of 4.5\,\mps.

\section{Stellar parameters}
\label{subsec-host-parameters}

\begin{table}
\caption{Stellar properties of G~9-40.
\label{table-C16_8748-stellar_parameters}}
\begin{center}
\begin{tabular}{lcr}
\hline
\hline
\noalign{\smallskip}
Parameter & Value & Source\\
\noalign{\smallskip}
\hline
\noalign{\smallskip}
\multicolumn{3}{c}{\emph{Coordinates and Main Identifiers}}\\
\noalign{\smallskip}
Name            & G~9-40                & \citet{Giclas1971} \\
                & NLTT~20661            & \citet{Luyten1979nlcs.book.....L} \\
                & K2-313                & Stef20\\
$\alpha$        &  08:58:52.33          & \gaia\ EDR3\\
$\delta$        & +21:04:34.2           & \gaia\ EDR3\\
EPIC ID         & 212048748             & EPIC\\
TIC ID          & 203214081             & TIC\\
Karmn ID        & J08588+210            & AF15\\
\noalign{\smallskip}
\multicolumn{3}{c}{\emph{Magnitudes}}\\
\noalign{\smallskip}
$V$ (mag)       & 13.82\,$\pm$\,0.04        & UCAC4\\
$R$ (mag)       & 12.68\,$\pm$\,0.07        & UCAC4\\
$G$ (mag)       & 12.7160\,$\pm$\,0.0028    & \gaia\ EDR3\\
$J$ (mag)       & 10.06\,$\pm$\,0.02        & 2MASS\\
$H$ (mag)       &  9.43\,$\pm$\,0.02        & 2MASS\\
$K$ (mag)       &  9.19\,$\pm$\,0.02        & 2MASS\\
\noalign{\smallskip}
\multicolumn{3}{c}{\emph{Parallax and kinematics}}\\
\noalign{\smallskip}
\splx{} (mas)                   & $35.930\pm0.025$      & \gaia\ EDR3\\
\sdist{} (pc)                   & $27.832\pm0.019$      & \gaia\ EDR3\\
\spmras{} (\maspyr)             & $+175.740\pm0.027$    & \gaia\ EDR3\\
\spmdec{} (\maspyr)             & $-318.332\pm0.020$    & \gaia\ EDR3\\ 
$U$ (\kmps)                     & $-$7.78\,$\pm$\,0.06    & This work\\
$V$ (\kmps)                     & $-$5.98\,$\pm$\,0.84    & This work\\
$W$ (\kmps)                     & $-$9.18\,$\pm$\,0.71    & This work\\
\noalign{\smallskip}
\multicolumn{3}{c}{\emph{Photospheric parameters}}\\
\noalign{\smallskip}
\steff{} (K)        & \steffv[]     & This work\\%
\logsg{} (dex)      & \logsgv[]     & This work\\%
\sfeh{} (dex)       & \sfehv[]      & This work\\%
\noalign{\smallskip}
\multicolumn{3}{c}{\emph{Physical parameters}}\\
\noalign{\smallskip}
\sm{} (\Msun)                   & \smv[]            & This work\\%
\sr{} (\Rsun)                   & \srv[]            & This work\\%
\slum{} ($10^{-4}$\,\Lsun)                 & \slumv[]          & This work\\%
$P_{\rm rot}$ (d)               & $30 \pm 1$        & Stef20\\%
\noalign{\smallskip}
\hline
\end{tabular}
\tablebib{
    Stef20: \citet{Stefansson2020AJ....159..100S};
    {\it Gaia} EDR3: \citet{GaiaEDR3};
    EPIC: \citet{EPIC};
    TIC: \citet{2018AJ....156..102S};
    AF15: \citet{AF15};
    UCAC4: \citet{UCAC4};
    2MASS: \citet{2MASS}.
}
\end{center}
\end{table}

%
%

We estimated stellar parameters of \host{} using the CARMENES-VIS co-added stellar template produced with \texttt{serval} and corrected for telluric features following \citet{Passegger2018A&A...615A...6P,Passegger2019A&A...627A.161P}. Using the measured upper limit $v\sin i$\,=\,2.0\,\kmps{}, we determined \steff\,=\,$3395\pm51\,\mathrm{K}$, \logsg\,=\,$4.84\pm0.04\,\mathrm{dex}$, and iron abundance \sfeh\,=\,$-0.07\pm0.16\,\mathrm{dex}$ of the star via spectral fitting with a grid of PHOENIX-SESAM models \citep{Passegger2019A&A...627A.161P}. We calculated the luminosity of G~9-40 by integrating the spectral energy distribution produced from multi-wavelength broadband photometric data as in \citet{Cifuentes2020A&A...642A.115C}. Using the Stefan–Boltzmann law, we determined the radius of the star and, with the mass-radius relationship from \citet{Schweitzer2019A&A...625A..68S}, the stellar mass. 

As a consistency check and for comparison with \citet{Stefansson2020AJ....159..100S}, we also analyzed the co-added CARMENES-VIS spectra using the {\tt SpecMatch-emp} software package \citep{SpecMatchemp}. {\tt SpecMatch-emp} estimates the stellar effective temperature \steff, radius \sr, and iron abundance \sfeh{} by fitting the spectral region between 5000\,{\AA} and 5900\,{\AA} to hundreds of library spectra gathered by the California Planet Search program. We found \steff\,=\,3400\,$\pm$\,70\,K, \sr\,=\,0.317\,$\pm$\,0.009\,\Rsun, and \sfeh\,=\,$-$0.15\,$\pm$\,0.12\,dex. Using the empirical equations by \citet{Torres2010A&ARv..18...67T} alongside \steff, \sfeh, and \sr{} above, we estimated the stellar mass to be \sm\,=\,0.2935\,$\pm$\,0.0072\,\Msun. 

The results from both methods using the CARMENES high-resolution spectra are in agreement within 1$\sigma$. Besides, they also agree with the stellar parameters presented in \citet{Stefansson2020AJ....159..100S} using the NIR high-resolution spectra of the star obtained with HPF. A summary of the stellar properties and other relevant parameters used in the remainder of this work is shown in Table~\ref{table-C16_8748-stellar_parameters}.



\section{Analysis and results}
\label{sec-analysis_and_results}

\begin{figure}
    \centering
    \includegraphics[width=\hsize]{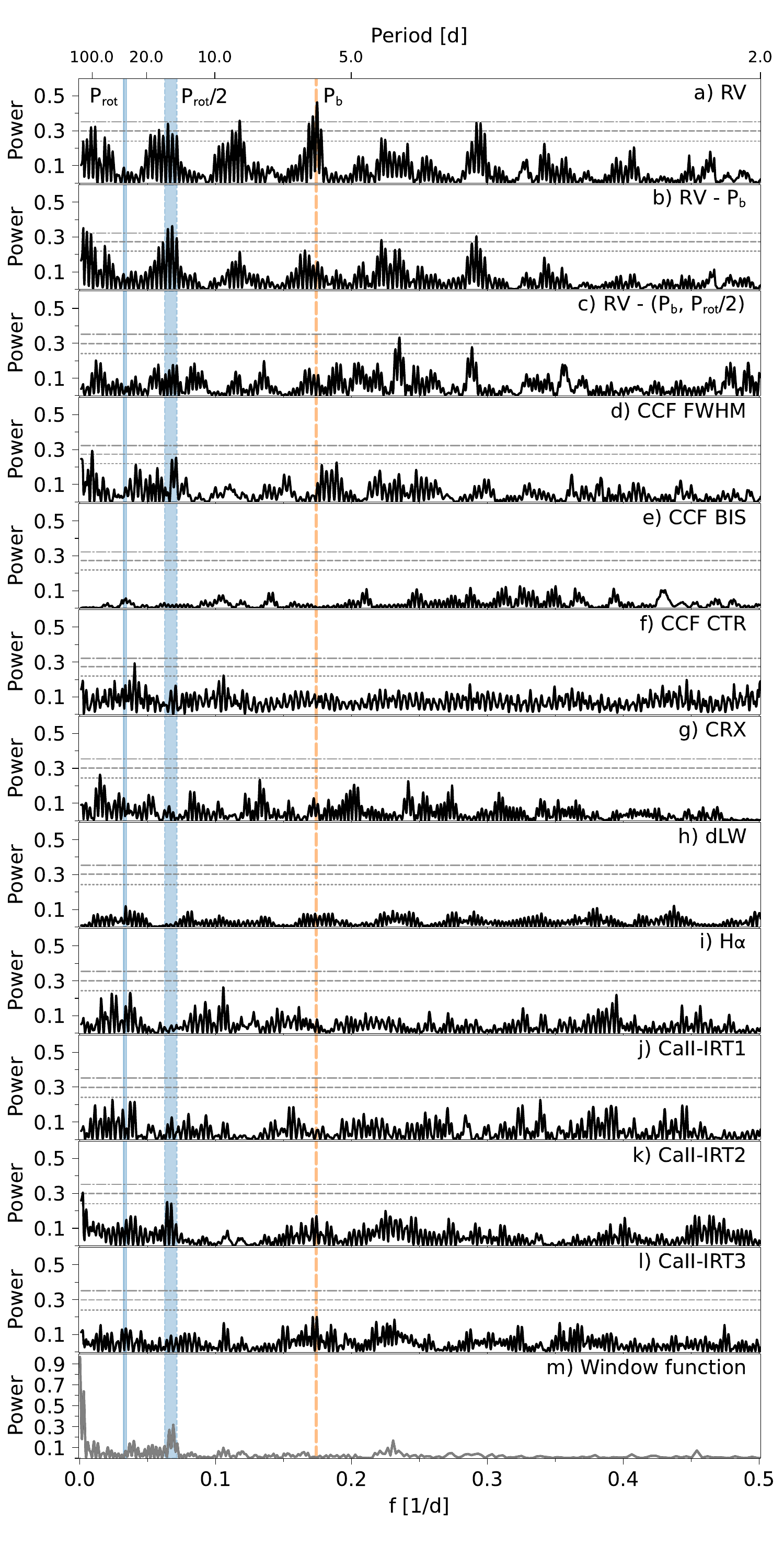}
    \caption{Generalized Lomb-Scargle periodograms of the CARMENES RVs and spectral activity indicators from \texttt{serval} and \texttt{raccoon}. Horizontal lines show the theoretical FAP levels of 10\% (short-dashed line), 1\% (long-dashed line), and 0.1\% (short-long-dashed line) for each panel. The vertical dashed lines mark the orbital frequencies of the transiting planet (${\rm f}_\mathrm{b}=0.174\,{\rm d}^{-1}$) and of the stellar signal at $0.033\,{\rm d}^{-1}$ and its second harmonic ($0.066\,{\rm d}^{-1}$) determined by \citet{Stefansson2020AJ....159..100S}. \emph{Panel a}: CARMENES RVs from \texttt{serval}. \emph{Panel b}: CARMENES RVs after modeling the signal from the transiting planet with a sinusoid. \emph{Panel c}: CARMENES RVs after modeling the signals from the transiting planet and the second harmonic of the rotation period of the star with a double sinusoid. \emph{Panels d--f}: CCF FWHM, CCF bisector span (CCF BIS), and CCF contrast (CCF CTR) from \texttt{raccoon}. \emph{Panels g-i}: Chromatic index (CRX), differential line width (dLW), H$\alpha$, and CaII IRT line indices from \texttt{serval}. \emph{Panel m}: Window function.}
    \label{fig:gls_rv}
\end{figure}

\begin{table}
    \centering
    \caption{Model comparison of RV-only fits with \texttt{juliet}. The winning model used for the joint fit is marked in boldface. }  \label{tab:models}
    \begin{tabular}{lccc}
        \hline
        \hline
        \noalign{\smallskip}
        Model   & GP kernel             & $\ln \mathcal{Z}$   & $K$ (m\,s$^{-1}$)   \\
        \noalign{\smallskip}
        \hline
 \noalign{\smallskip}
 \multicolumn{4}{c}{\textit{Base models}}\\
 \noalign{\smallskip}
0p              & \dots                 & $-$302.8    & \dots    \\
0p+GP1          & EXP\tablefootmark{a}  & $-$286.3    & \dots    \\
0p+GP2          & ESS\tablefootmark{b}  & $-$285.8    & \dots    \\
 \noalign{\smallskip}
 \multicolumn{4}{c}{\textit{One-signal models}}\\
 \noalign{\smallskip}
1p              & \dots                 & $-$298.4    & $2.89\pm0.75$    \\
 \noalign{\smallskip}
 \multicolumn{4}{c}{\textit{Two-signal models}}\\
 \noalign{\smallskip}
1p+Sin          & \dots                 & $-$288.0    & $3.67\pm0.63$    \\
1p+GP1          & EXP\tablefootmark{a}  & $-$282.4    & $3.17\pm0.80$    \\
{\bf 1p+GP2}          & {\bf ESS\tablefootmark{b}}  & {\bf -279.2}    & $3.27\pm0.52$    \\
 \noalign{\smallskip}
 \multicolumn{4}{c}{\textit{Three-signal models}}\\
 \noalign{\smallskip}
2p+Sin          & \dots                 & $-$291.6    & $3.23^{+0.49}_{-0.41}$    \\
                & &                                 & $2.00\pm0.46$    \\
2p+GP1          & EXP\tablefootmark{a}  & $-$284.6    & $2.75^{+0.61}_{-0.50}$    \\
                & &                                 & $2.52^{+0.42}_{-0.55}$    \\
2p+GP2          & ESS\tablefootmark{b}  & $-$281.6    & $3.23^{+0.46}_{-0.48}$    \\
                & &                                 & $2.19^{+0.51}_{-0.61}$    \\
        \noalign{\smallskip}
        \hline
    \end{tabular}
    \tablefoot{
        \tablefoottext{a}{Squared-exponential kernel of the form $k_{i,j} = \sigma^2_\mathrm{GP,RV} \exp\left(-|t_i - t_j|/T_\mathrm{GP,RV}\right)$}.
        \tablefoottext{b}{Exponential-sine-squared kernel of the form $k_{i,j} = \sigma^2_\mathrm{GP,RV} \exp\left(- \alpha_\mathrm{GP,RV} (t_i - t_j)^2 - \Gamma_\mathrm{GP,RV} \sin^2 \left[\frac{\pi |t_i - t_j|}{P_{\rm rot;GP,RV}}\right]\right)$.}
    }
\end{table}

For the modeling of the photometric and RV data in this work, we used \texttt{juliet} \citep{juliet}, a \texttt{Python}-based fitting package that applies nested samplers to explore efficiently the parameter space of a given prior volume and compute the Bayesian model log evidence, $\ln \mathcal{Z}$. We used \texttt{dynesty} \citep{dynesty} as our dynamic nested sampling algorithm. The \texttt{juliet} package is built on many publicly available tools for the modeling of transits \citep[\texttt{batman},][]{batman}, RVs \citep[\texttt{radvel},][]{radvel}, and GPs (\texttt{george}, \citealt{Ambikasaran2015ITPAM..38..252A}; \texttt{celerite}, \citealt{celerite}). Due to its efficient computation of the $\ln \mathcal{Z}$, we can compare statistically models with different numbers of parameters accounting for the model complexity and the number of degrees of freedom. We consider a model to be moderately favored over another if the difference in its Bayesian log evidence is greater than two, and strongly favored if it is greater than five \citep{2008ConPh..49...71T}. If $\Delta \ln \mathcal{Z} \lesssim 2$, then the models are indistinguishable so the simpler model with fewer degrees of freedom would be chosen. 

\subsection{Photometric fit}
\label{subsec-phot_only}

\begin{figure*}[th]
    \centering
    \includegraphics[width=0.41\hsize]{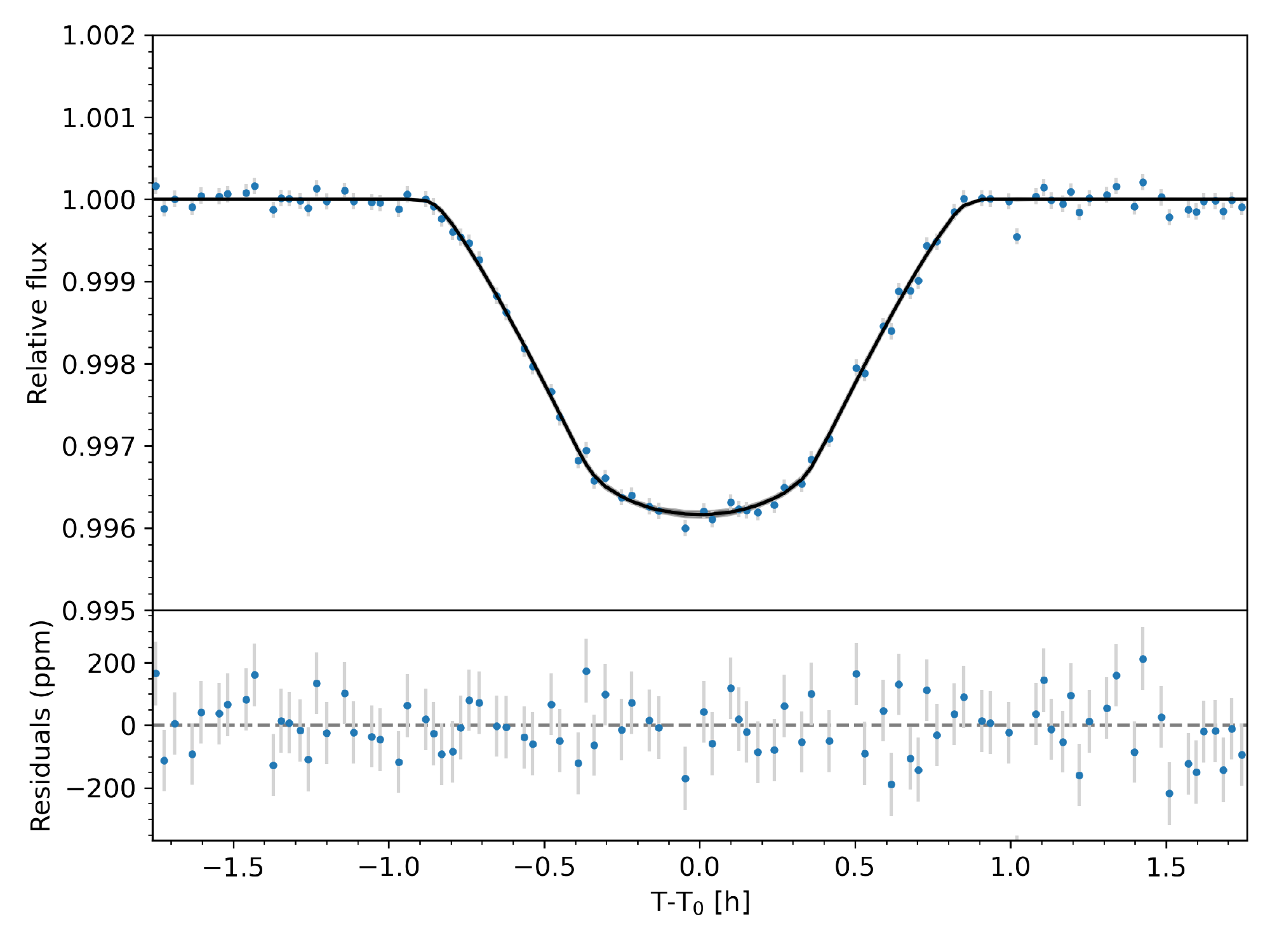}
    \includegraphics[width=0.41\hsize]{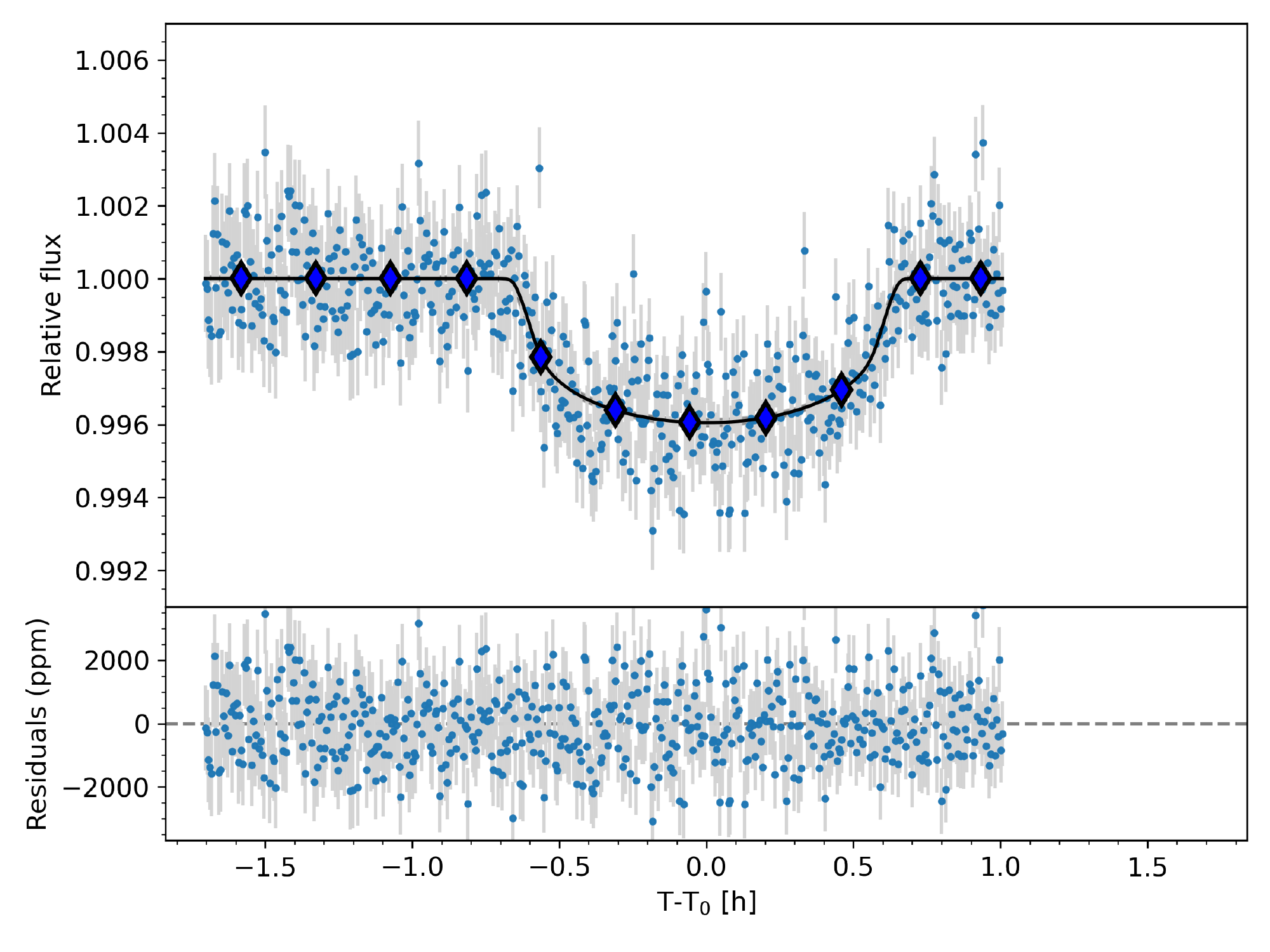}\\
    \includegraphics[width=0.41\hsize]{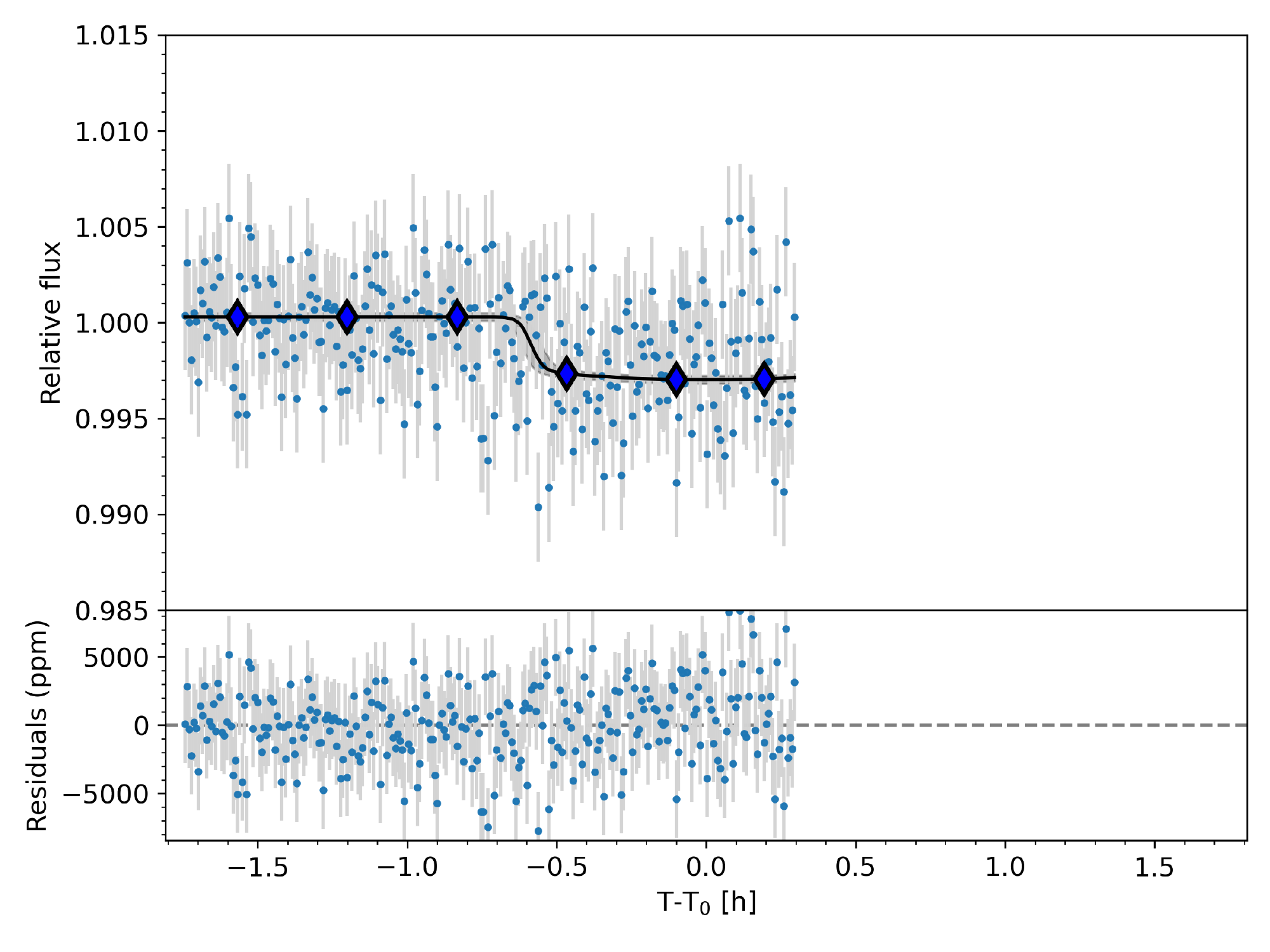}
    \includegraphics[width=0.41\hsize]{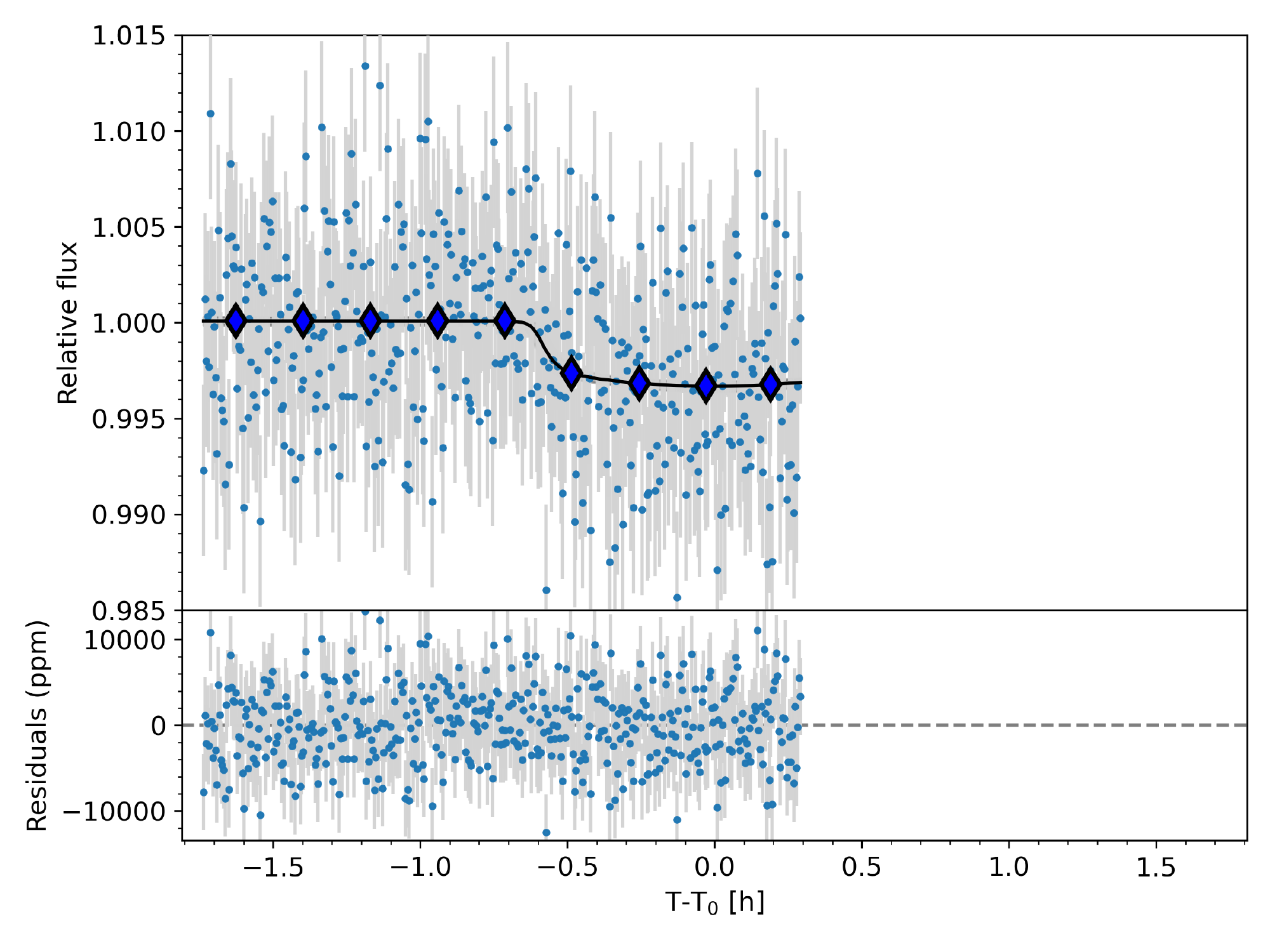}\\
    \includegraphics[width=0.41\hsize]{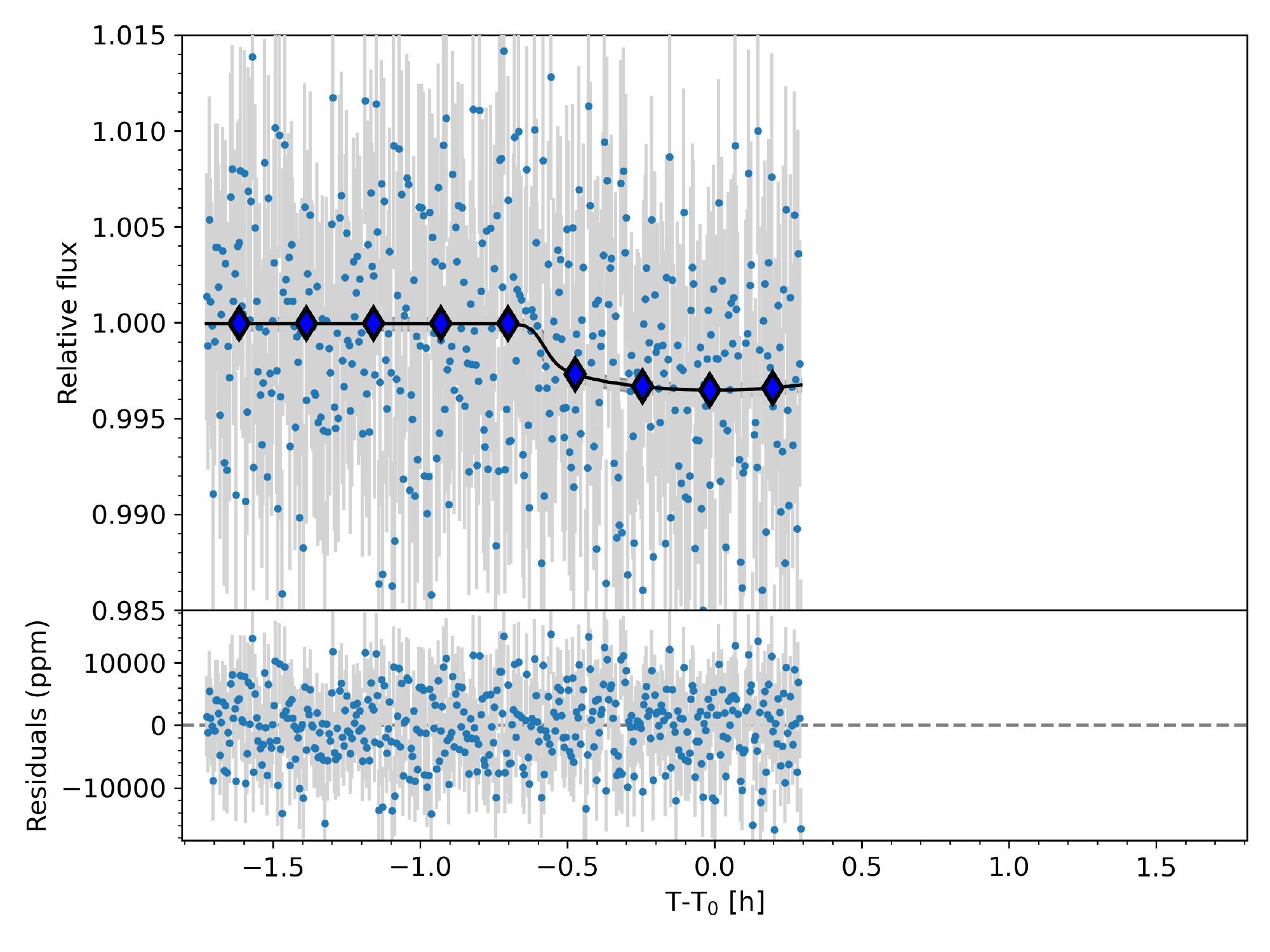}
    \includegraphics[width=0.41\hsize]{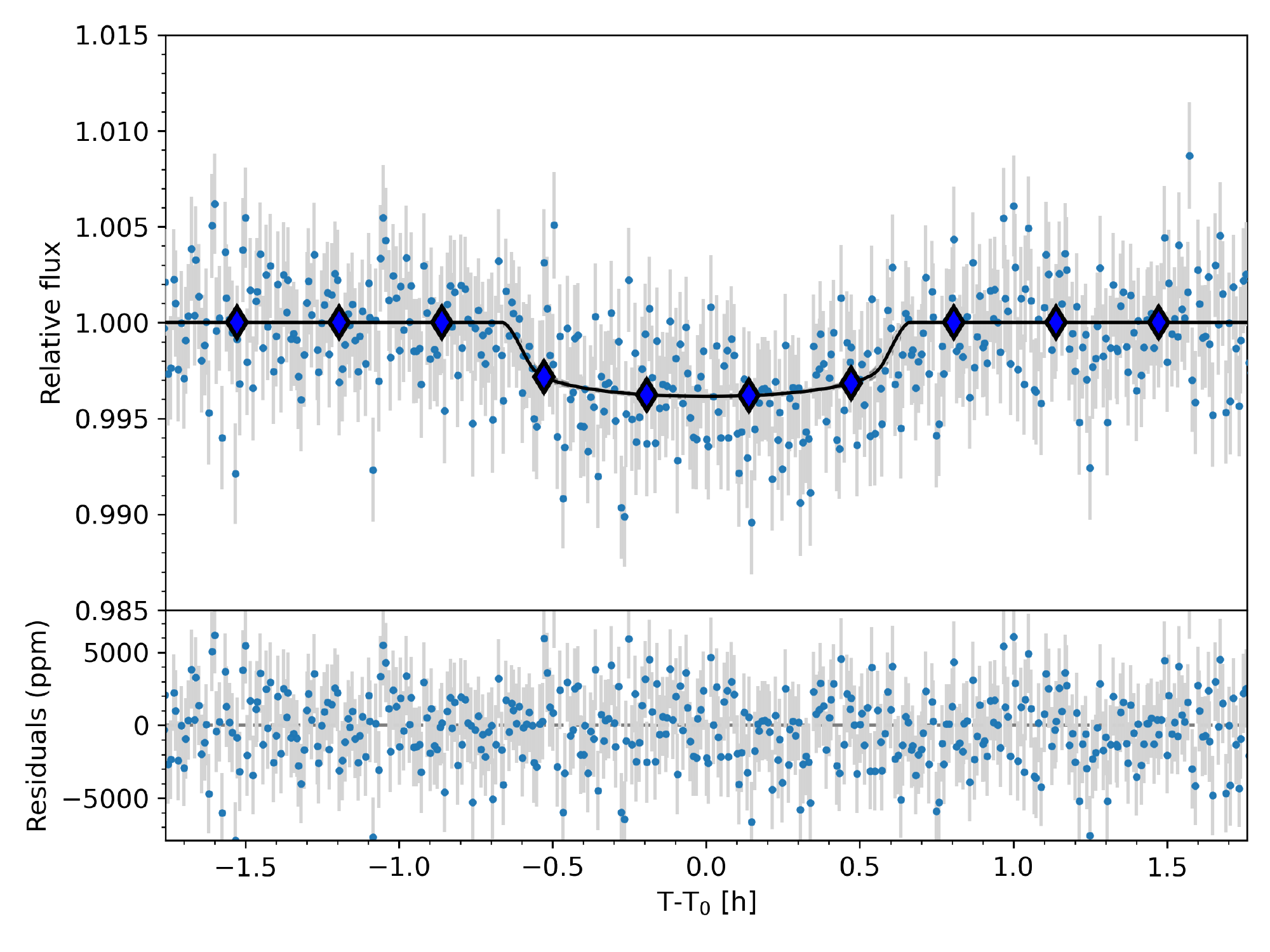}\\
    \includegraphics[width=0.41\hsize]{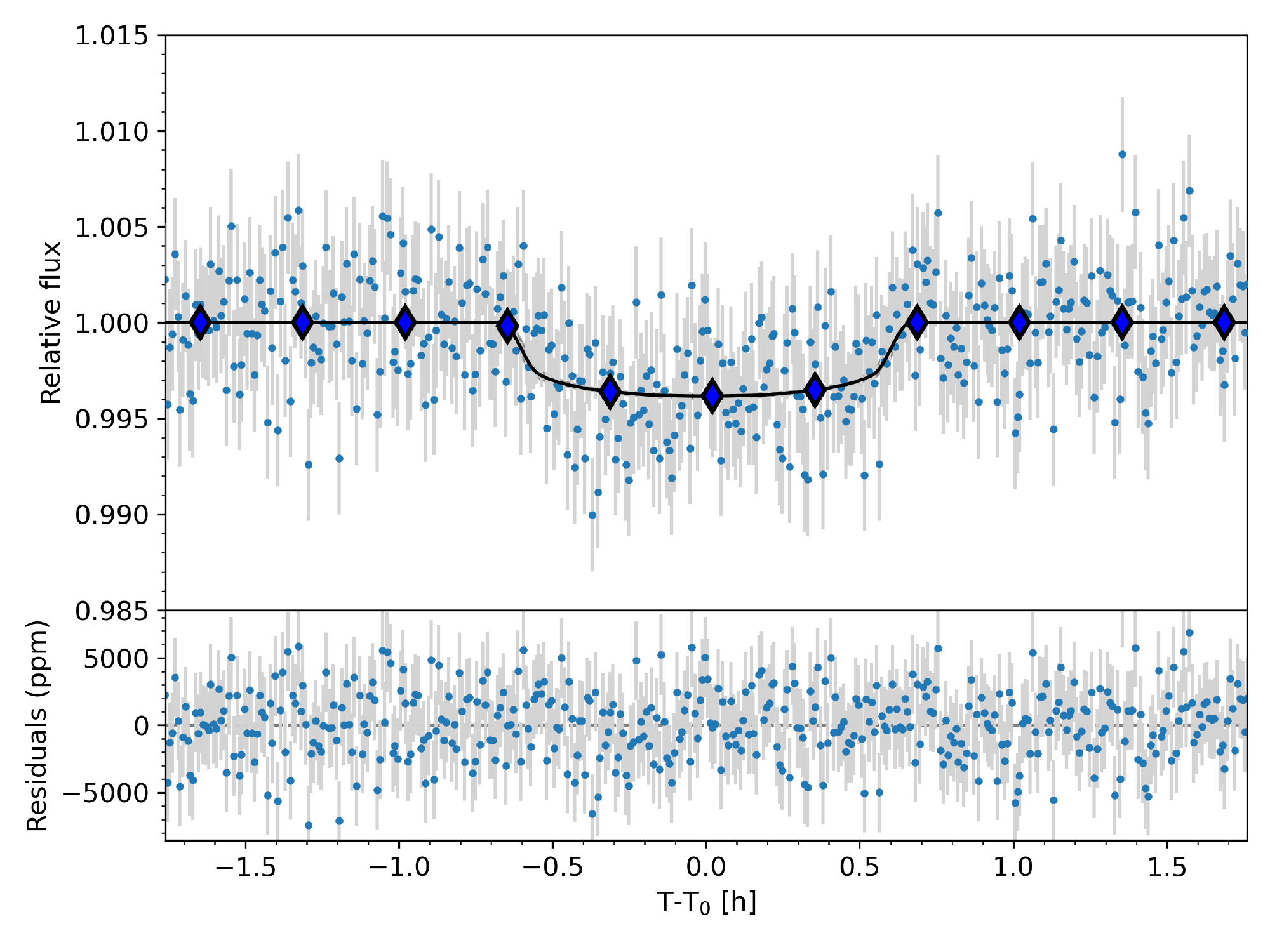}
    \includegraphics[width=0.41\hsize]{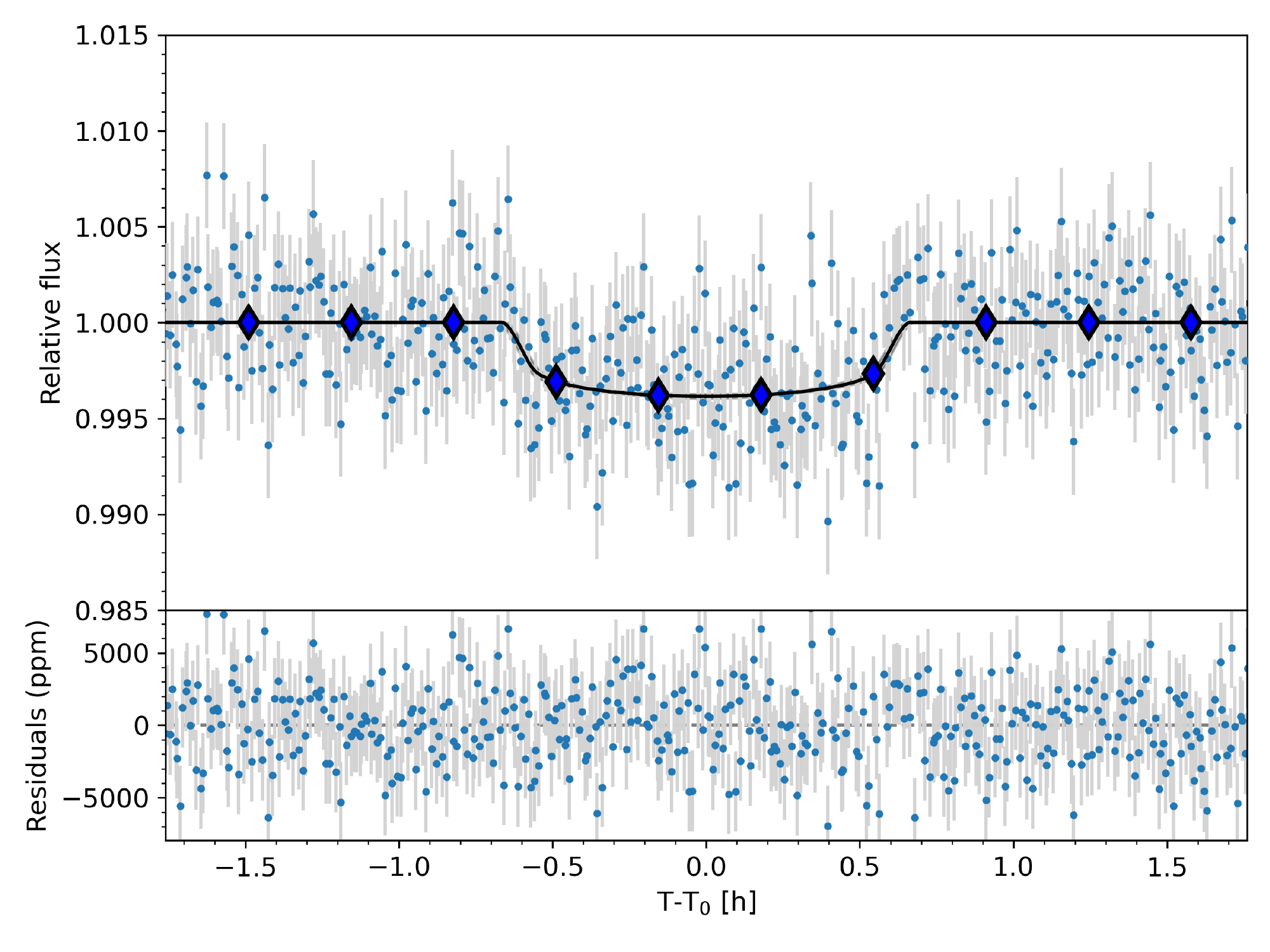}
    \caption{Phase-folded light curves of \planetb{}. {\textit Top row}: transits observed with \textit{K2} (left) and ARCTIC (right). {\textit Second row}: transits observed with MuSCAT2 in $r$-band (left) and $i$-band (right). {\textit Third row}: transits observed with MuSCAT2 in $z_s$-band (left) and \textit{TESS} Sector 44 (right). \textit{Bottom row}: transits observed with \textit{TESS} in Sectors 45 (left) and 46 (right). In all panels, the black lines and shaded areas indicate the detrended best fit model from Sect.~\ref{subsec-analysis-joint_model} and its 1$\sigma$ confidence interval, respectively. Below each panel, the residuals after the subtraction of the median best fit model are shown. Blue diamonds show binned points to improve visualization.
    }
    \label{fig:k2_lc_p1}
\end{figure*}

To update and improve the ephemeris of \planetb{} since its initial discovery, we first model all the available transit photometry using \texttt{juliet}. For the transiting planet, we followed the $r_1$, $r_2$ mathematical parametrization introduced by \citet{Espinoza18} to fit the planet-to-star radius ratio $p=R_p/R_*$ and the impact parameter of the orbit $b$. We fit the stellar density ($\rho_\star$) rather than the scaled semi-major axis ($a/R_\star$) using an informed normal prior based on the stellar parameters from Table~\ref{table-C16_8748-stellar_parameters}. The \textit{TESS} and \textit{K2} data are modeled with a quadratic limb darkening law (shared among sectors for \textit{TESS}), while the ground-based data were modeled with linear limb darkening laws, both parametrized with the uniform sampling scheme of \citet{Kipping13}. Using a quadratic and a linear limb darkening law is the recommended approach by \citet{EspinozaJordan2016MNRAS.457.3573E} when dealing simultaneously with space- and ground-based photometry. Each instrument is modeled including an additional jitter term and fixing the dilution factor (i.e., the amount that a light curve is diluted due to neighboring stellar contamination) to one due to the absence of nearby companions as shown in Sect.~\ref{subsec-observations-hci}. Finally, to remove the stellar variability in the \textit{K2} light curve (Fig.~\ref{fig:tess_lc}), we used the exponential GP kernel from \texttt{celerite} \citep{celerite}
\begin{equation*}
    k_{i,j} = \sigma^2_\mathrm{GP,K2} \exp\left(- |t_i - t_j|/T_\mathrm{GP,K2}\right),
\end{equation*}
where $T_\mathrm{GP,K2}$ is the characteristic timescale in days and $\sigma_\mathrm{GP,TESS}$ the amplitude of the GP modulation in parts-per-million (ppm). While we used a GP kernel to remove the stellar variability in the \textit{K2} light curve, it was not necessary for the \textit{TESS} data (i.e., $\Delta \ln\mathcal{Z} < 2$ when including and not including the GP term). 

We used informed priors for the period ($P$) and the mid-transit time ($t_0$) for the transiting planet based on the results from \citet{Stefansson2020AJ....159..100S}. Our prior choices for the remaining parameters are the same as for the final joint fit shown in Table~\ref{tab:posteriors}. The posterior distributions for this photometry-only analysis are indistinguishable from the final joint fit including the RV data, so we do not discuss them in the text and directly refer to Tables~\ref{tab:posteriors} and \ref{tab:derivedparams}. By adding the MuSCAT2 and \textit{TESS} light curves we are able to reduce the uncertainty in the period of the transiting planet by a factor of 3. In addition, we further update and constrain the ephemeris of the planet to an uncertainty to within a few minutes for the next 5 years, coincident with the primary mission of the James Webb Space Telescope (\textit{JWST}). Besides, with the new data, we are able to reduce the uncertainty in the planet radius from 5.4\% \citep{Stefansson2020AJ....159..100S} to 3.4\% (this work, Table~\ref{tab:derivedparams}).

\subsection{Radial velocity fit}
\label{subsec-rv_only}

Before modeling the complete RV data set, we carried out an unbiased signal search in the spectroscopic data using generalized Lomb-Scargle (GLS) periodograms \citep{GLS}. In this way, we can inform our fits about how many signals are present in the data and choose the most appropriate model for each signal according to its origin. Figure~\ref{fig:gls_rv} shows the GLS of the CARMENES-only RV data and several activity indicators extracted with \texttt{serval} and \texttt{raccoon}. We initially do not include the IRD or HPF datasets because of their low number of measurements and inadequate cadence for this type of analysis. 

The highest peak in the GLS of the CARMENES RV data (Fig.~\ref{fig:gls_rv}a, ${\rm FAP} \ll 0.1\%$) coincides with the orbital frequency of \planetb{}, indicating that the planet is independently detected in the RV dataset alone. We subtract the Doppler signal of \planetb{} from the CARMENES RVs assuming a circular Keplerian model with the period and mid-transit time constrained from our previous photometry-only analysis (Sect.~\ref{subsec-phot_only}). The GLS of the residuals (Fig.~\ref{fig:gls_rv}b) show the highest power at the frequency of the second harmonic of the stellar rotational period at 15\,d, which is also seen (with lower significance) in the GLS of the CCF FWHM and the second line of the CaII~IRT. We do not see any power at the rotational period itself in the RV residuals or the activity indicators. After modeling this signal associated with the activity of the host star with a sinusoid simultaneously together with the transiting planet, we are left with a signal with an ${\rm FAP} \sim 0.1\%$ at around 3.8\,d that has no counterpart in other spectral indicators (Fig.~\ref{fig:gls_rv}c). 

To summarize the results from the frequency analysis using GLS periodograms, we find several signals in the CARMENES RV data with high statistical significance. The most significant signal in the CARMENES RV data is the one associated with the transiting planet at 5.75\,d. Furthermore, the RV residuals after modeling for this signal reveal additional signals at 15\,d and 3.8\,d. To understand and confirm the nature of these signals, we carry out a model comparison analysis on the RV data only. Here, we use the full RV dataset for completeness, and to extend the time baseline of the observations. The results comparing the Bayesian log evidences are summarized in Table~\ref{tab:models}.

The stellar rotation period of $30\pm1$\,d, as determined by \citet{Stefansson2020AJ....159..100S} for \host{} using \textit{K2} photometry, is not present in the RVs, but rather its second harmonic at 15\,d. The harmonics of the rotation period often appear in RV data because of various factors such as the longitudinal distribution of surface features or the stellar inclination \citep[e.g.,][]{Boisse2011A&A...528A...4B}. Some examples are GJ~514 \citep{Lafarga2021A&A...652A..28L}, where the rotational period of the star is determined from photometry and only the second harmonic of the rotational period shows a significant signal in the RVs probably due to two spots separated by 180\,deg in longitude; or GJ~649 \citep{Rosenthal2021ApJS..255....8R,Lafarga2021A&A...652A..28L}, where the RVs show a significant signal at half the rotational period, but the H$\alpha$ and CaII~IRT activity indices show signals at the true rotational period. 

Therefore, we attribute the 15\,d signal present in the RV data to stellar activity rather than a non-transiting planetary companion, which it is also confirmed by its presence in the CCF FWHM and CaII IRT activity indices (Figs.~\ref{fig:gls_rv}d and \ref{fig:gls_rv}k). To account for this signal, we try different models and study the goodness of the fit using the Bayesian log evidence $\ln \mathcal{Z}$. First, assuming that the surface pattern of the star has not changed over the time covered by the RV observations, we model the 15\,d signal with a sinusoid together with a Keplerian (either circular or eccentric) for the transiting planet (Model 1p+Sin in Table~\ref{tab:models}). Our results show that including the 15\,d signal improves the $\ln \mathcal{Z}$ ($\Delta \ln \mathcal{Z} > 10$) with respect to a model that includes only the transiting planet (Model 1p), no matter whether the transiting planet is in a circular orbit or an eccentric one. The eccentric models, despite having a higher nominal $\ln \mathcal{Z}$, are indistinguishable from the simpler circular models and provide values consistent with zero eccentricity at 1$\sigma$. 

The approximately 900\,d baseline of the RV observations is sufficiently long to cover many stellar rotations. Therefore, we tested whether a GP, rather than a simple sinusoid, would be more suitable to model the 15\,d signal given its stellar activity origin. Indeed, modeling the 15\,d signal using different GP kernels yields much better results compared to a sinusoid ($\Delta \ln \mathcal{Z} = \ln \mathcal{Z}_{\rm 1p+GP} - \ln \mathcal{Z}_{\rm 1p+Sin} > 5$). Modeling signals associated with stellar activity using GPs typically provide the best results in terms of log evidence if the cadence, baseline, and number of measurements is adequate (Stock et al. 2022, in prep.). We try two different GP kernels to model this effect: a squared-exponential kernel (EXP, GP1) and a quasi-periodic kernel (ESS, GP2), both from \texttt{george} \citep{Ambikasaran2015ITPAM..38..252A}. The model with the highest $\ln \mathcal{Z}$ is 1p+GP2, which represents a circular orbit for the transiting planet and a quasi-periodic ESS kernel to account for the stellar activity effects imprinted on the RV data. This model is able to represent the 15\,d variability in the RV data better ($\Delta \ln \mathcal{Z} > 3$) than the other kernel while obtaining the smallest uncertainty in the RV semi-amplitude of the transiting planet. We tried different priors for the $P_{\rm rot}$ hyperparameter in the ESS kernel, but the results were very similar in terms of $\ln \mathcal{Z}$ and the mass determination of the transiting planet. Wide uniform priors from 10 to 50 days provide the same results as the final ones presented in Table~\ref{tab:models}, with the ESS finding a value of $P_{\rm rot} \sim 15$\,d. We tried a narrow prior centered around 30\,d in the ESS kernel, but the results were inconclusive and worse than a wide or narrow prior that include the 15\,d region. This makes sense considering that the 30\,d signal itself is small in the GLS periodograms (see Fig.~\ref{fig:gls_rv}).

To check the significance of the signal associated with the transiting planet and to confirm that the nature of the 15\,d signal is of non-planetary origin we carried out another set of models. It is interesting to compare a model including a GP component only (0p+GP) with one that also includes a Keplerian orbit. GP models can act as a high-pass filter that removes all the variability in the RV data, absorbing planetary signals that are at the level of the instrumental precision. However, we find that models accounting for the transiting planet are favored ($\Delta \ln \mathcal{Z} = \ln \mathcal{Z}_{\rm 1p+GP} - \ln \mathcal{Z}_{\rm 0p+GP} > 5$) over GP-only models, which corroborates our detection and the conjecture that the RV data alone would have been able to detect the transiting planet independently.

Finally, we checked if the signal at around 3.8\,d visible in Fig.~\ref{fig:gls_rv}c is still significant after modeling the 15\,d signal not with a sinusoid, but with a quasi-periodic GP kernel. We find that adding an extra circular Keplerian to account for the 3.8\,d signal (2p) to the previous models (1p+Sin, 1p+GP1, and 1p+GP2) makes the fit less likely in all cases, although the statistical difference is only moderately significant ($\Delta \ln \mathcal{Z} < 2-3$). 
Therefore, we do not model the 3.8\,d signal in our final fit, which includes a circular orbit for the transiting planet and a quasi-periodic kernel that accounts for the signal at the second harmonic of the stellar rotation (1p+GP2).


\subsection{Joint modeling of the light curves and RVs}
\label{subsec-analysis-joint_model}

\begin{table}[t]
    \centering
    {\footnotesize
    \caption{Priors, median and 68\% credibility intervals of the posterior distributions for each fit parameter of the final joint model obtained for the \host{} system using \texttt{juliet}.}
    \label{tab:posteriors}
    \begin{tabular}{l@{\hspace{1mm}}l@{\hspace{3mm}}r@{\hspace{3mm}}}
        \hline
        \hline
        \noalign{\smallskip}
        Parameter & Prior & Posterior  \\
        \noalign{\smallskip}
        \hline
        \noalign{\smallskip}
        \multicolumn{3}{c}{\it Stellar parameters} \\[0.1cm]
        \noalign{\smallskip}
        $\rho_\star$ ($\mathrm{g\,cm\,^{-3}}$)  & $\mathcal{N}(1.5,0.15)$     & $1.62\pm0.11$ \\[0.1 cm]
        \noalign{\smallskip}
        \multicolumn{3}{c}{\it Orbit parameters} \\[0.1cm]
        \noalign{\smallskip}                                             
        $P$ (d)                     & $\mathcal{N}(5.746,0.050)$    & $5.7459982 \pm 0.0000020$  \\[0.1 cm]
        $t_0$\tablefootmark{(a)}    & $\mathcal{N}(9503.32,0.50)$   & $9503.32682 \pm 0.00042$  \\[0.1 cm]
        $r_1$                       & $\mathcal{U}(0,1)$            & $0.675 \pm 0.028$  \\[0.1 cm]
        $r_2$                       & $\mathcal{U}(0,1)$            & $0.05750 \pm 0.00075$  \\[0.1 cm]
        $e$                         & Fixed                         & $0.0$  \\[0.1 cm]
        $\omega$ (deg)              & Fixed                         & $90.0$  \\[0.1 cm]
        $K$ ($\mathrm{m\,s^{-1}}$)  & $\mathcal{U}(0,20)$           & $3.22\pm0.49$  \\
        \noalign{\smallskip}
        \multicolumn{3}{c}{\it Photometry parameters} \\[0.1cm]
        \noalign{\smallskip}
        $\sigma_{\textnormal{K2}}$ (ppm)        & $\mathcal{J}(1,10^4)$         & $96.4\pm2.8$ \\[0.1 cm] 
        $q_{1,\textnormal{K2}}$                 & $\mathcal{U}(0,1)$            & $0.84^{+0.10}_{-0.17}$ \\[0.1 cm]
        $q_{2,\textnormal{K2}}$                 & $\mathcal{U}(0,1)$            & $0.18^{+0.15}_{-0.11}$ \\[0.1 cm]
        $\sigma_{\textnormal{TESS,S44}}$ (ppm)  & $\mathcal{J}(1,10^4)$         & $10.0^{+36.7}_{-7.7}$ \\[0.1 cm] 
        $\sigma_{\textnormal{TESS,S45}}$ (ppm)  & $\mathcal{J}(1,10^4)$         & $8.0^{+30.0}_{-5.9}$ \\[0.1 cm] 
        $\sigma_{\textnormal{TESS,S46}}$ (ppm)  & $\mathcal{J}(1,10^4)$         & $8.6^{+28.4}_{-6.4}$ \\[0.1 cm] 
        $q_{1,\textnormal{TESS}}$               & $\mathcal{U}(0,1)$            & $0.59^{+0.24}_{-0.25}$ \\[0.1 cm]
        $q_{2,\textnormal{TESS}}$               & $\mathcal{U}(0,1)$            & $0.22^{+0.24}_{-0.15}$ \\[0.1 cm]
        $\sigma_{\textnormal{M2r}}$ (ppm)       & $\mathcal{J}(10^{-2},10^5)$   & $4.3^{+150}_{-4.2}$ \\[0.1 cm] 
        $M_{\textnormal{M2r}}$ (ppm)            & $\mathcal{U}(-0.1,0.1)$       & $-0.00029\pm0.00018$ \\[0.1 cm]
        $q_{1,\textnormal{M2r}}$                & $\mathcal{U}(0,1)$            & $0.23^{+0.24}_{-0.15}$ \\[0.1 cm]
        $\sigma_{\textnormal{M2i}}$ (ppm)       & $\mathcal{J}(10^{-2},10^5)$   & $5.7^{+220}_{-5.6}$ \\[0.1 cm] 
        $M_{\textnormal{M2i}}$ (ppm)            & $\mathcal{U}(-0.1,0.1)$       & $-0.00008\pm0.00022$ \\[0.1 cm]
        $q_{1,\textnormal{M2i}}$                & $\mathcal{U}(0,1)$            & $0.52^{+0.28}_{-0.30}$ \\[0.1 cm]
        $\sigma_{\textnormal{M2z}}$ (ppm)       & $\mathcal{J}(10^{-2},10^5)$   & $4.5^{+170}_{-4.5}$ \\[0.1 cm] 
        $M_{\textnormal{M2z}}$ (ppm)            & $\mathcal{U}(-0.1,0.1)$       & $-0.00008\pm0.00032$ \\[0.1 cm]
        $q_{1,\textnormal{M2z}}$                & $\mathcal{U}(0,1)$            & $0.65^{+0.23}_{-0.33}$ \\[0.1 cm]
        $\sigma_{\textnormal{ARCTIC}}$ (ppm)    & $\mathcal{J}(10^{-2},10^2)$   & $1.5^{+23.3}_{-1.4}$ \\[0.1 cm] 
        $M_{\textnormal{ARCTIC}}$ (ppm)         & $\mathcal{U}(-0.1,0.1)$       & $0.000007\pm0.000048$ \\[0.1 cm]
        $q_{1,\textnormal{ARCTIC}}$             & $\mathcal{U}(0,1)$            & $0.78\pm0.08$ \\[0.1 cm]
        \noalign{\smallskip}
        \multicolumn{3}{c}{\it RV parameters} \\[0.1cm]
        \noalign{\smallskip}
        $\gamma_{\textnormal{CARM}}$ ($\mathrm{m\,s^{-1}}$)        & $\mathcal{U}(-100,100)$   & $2.90^{+3.40}_{-2.71}$ \\[0.1 cm]
        $\sigma_{\textnormal{CARM}}$ ($\mathrm{m\,s^{-1}}$)     & $\mathcal{J}(0.1,100)$    & $0.75^{+0.72}_{-0.51}$ \\[0.1 cm]
        $\gamma_{\textnormal{IRD}}$ ($\mathrm{m\,s^{-1}}$)         & $\mathcal{U}(-100,100)$   & $4.40^{+3.08}_{-3.15}$ \\[0.1 cm]
        $\sigma_{\textnormal{IRD}}$ ($\mathrm{m\,s^{-1}}$)      & $\mathcal{J}(0.1,100)$    & $1.01^{+2.48}_{-0.76}$ \\[0.1 cm]
        $\gamma_{\textnormal{HPF}}$ ($\mathrm{m\,s^{-1}}$)         & $\mathcal{U}(-100,100)$   & $0.29^{+5.96}_{-5.98}$ \\[0.1 cm]
        $\sigma_{\textnormal{HPF}}$ ($\mathrm{m\,s^{-1}}$)      & $\mathcal{J}(0.1,100)$    & $0.90^{+2.78}_{-0.67}$ \\
        \noalign{\smallskip}
        \multicolumn{3}{c}{\it GP hyperparameters} \\
        \noalign{\smallskip}
        $\sigma_\mathrm{GP,K2}$ (ppm)                   & $\mathcal{J}(10^{-10},10^{-2})$   & $450^{+140}_{-30}$ \\[0.1 cm]
        $T_\mathrm{GP,K2}$ (d)                          & $\mathcal{J}(10^{-6},10^{-2})$    & $0.0002^{+0.0004}_{-0.0001}$ \\[0.1 cm]
        $\sigma_\mathrm{GP,RV}$ ($\mathrm{m\,s^{-1}}$)  & $\mathcal{J}(0.01,30)$            & $6.54^{+2.46}_{-1.54}$ \\[0.1 cm]
        $\alpha_\mathrm{GP,RV}$ (d$^{-2}$)              & $\mathcal{J}(10^{-8},10^{-2})$    & $0.00035^{+0.00042}_{-0.00020}$ \\[0.1 cm]
        $\Gamma_\mathrm{GP,RV}$                         & $\mathcal{J}(10^{-2},10^1)$       & $1.00^{+2.52}_{-0.64}$ \\[0.1 cm]
        $P_\mathrm{rot;GP,RV}$ (d)                      & $\mathcal{N}(15,2)$               & $14.46^{+0.42}_{-0.71}$ \\
        \noalign{\smallskip}
        \hline
    \end{tabular}
    \tablefoot{
        \tablefoottext{a}{Units are BJD - 2450000.}
        The prior labels of $\mathcal{N}$, $\mathcal{U}$, and $\mathcal{J}$ represent normal, uniform, and Jeffrey's distributions, respectively.
    }
    }
\end{table}

\begin{table}[t]
    \centering
    \caption{Derived planetary parameters obtained for the \host{} system using the posterior values from Table~\ref{tab:posteriors} and stellar parameters from Table~\ref{table-C16_8748-stellar_parameters}.}
    \label{tab:derivedparams}
    \begin{tabular}{lr} 
        \hline
        \hline
        \noalign{\smallskip}
        Parameter\tablefootmark{(a)} & \planetb{}  \\
        \noalign{\smallskip}
        \hline
        \noalign{\smallskip}
        \multicolumn{2}{c}{\it Derived transit parameters} \\[0.1cm]
        \noalign{\smallskip}
        $p = R_{\rm p}/R_\star$             & $0.05750 \pm 0.00075$  \\[0.1 cm]
        $b = (a/R_\star)\cos i_{\rm p}$     & $0.513\pm0.042$ \\[0.1 cm]
        $a/R_\star$                         & $30.46^{+0.67}_{-0.75}$  \\[0.1 cm]
        $i_p$ (deg)                         & $89.03 \pm 0.10$  \\[0.1 cm]
        $t_T$ (h)                           & $1.693^{+0.070}_{-0.061}$  \\[0.1 cm]
        \noalign{\smallskip}
        \multicolumn{2}{c}{\it Derived physical parameters} \\[0.1cm]
        \noalign{\smallskip}
        $M_{\rm p}$ ($M_\oplus$)            & $4.00 \pm 0.63$ \\[0.1 cm]
        $R_{\rm p}$ ($R_\oplus$)            & $1.900 \pm 0.065$  \\[0.1 cm]
        $\rho_{\rm p}$ (g cm$^{-3}$)        & $3.20^{+0.63}_{-0.58}$  \\[0.1 cm]
        $g_{\rm p}$ (cm s$^{-2}$)           & $1080 \pm 180$  \\[0.1 cm]
        $a_{\rm p}$ (au)                    & $0.04180 \pm 0.00064$ \\[0.1 cm]
        $S$ ($S_\oplus$)                    & $6.27 \pm 0.19$  \\[0.1 cm]
        $T_\textnormal{eq}$ (K)\tablefootmark{(b)}      & $440.6 \pm 7.6$ \\[0.1 cm]
        \noalign{\smallskip}
        \hline
    \end{tabular}
    \tablefoot{
      \tablefoottext{a}{Error bars denote the $68\%$ posterior credibility intervals.}
      \tablefoottext{b}{Equilibrium temperatures were calculated assuming zero Bond albedo and uniform surface temperature across the entire planet.}
      }
\end{table}

\begin{figure}
    \centering
    \includegraphics[width=\hsize]{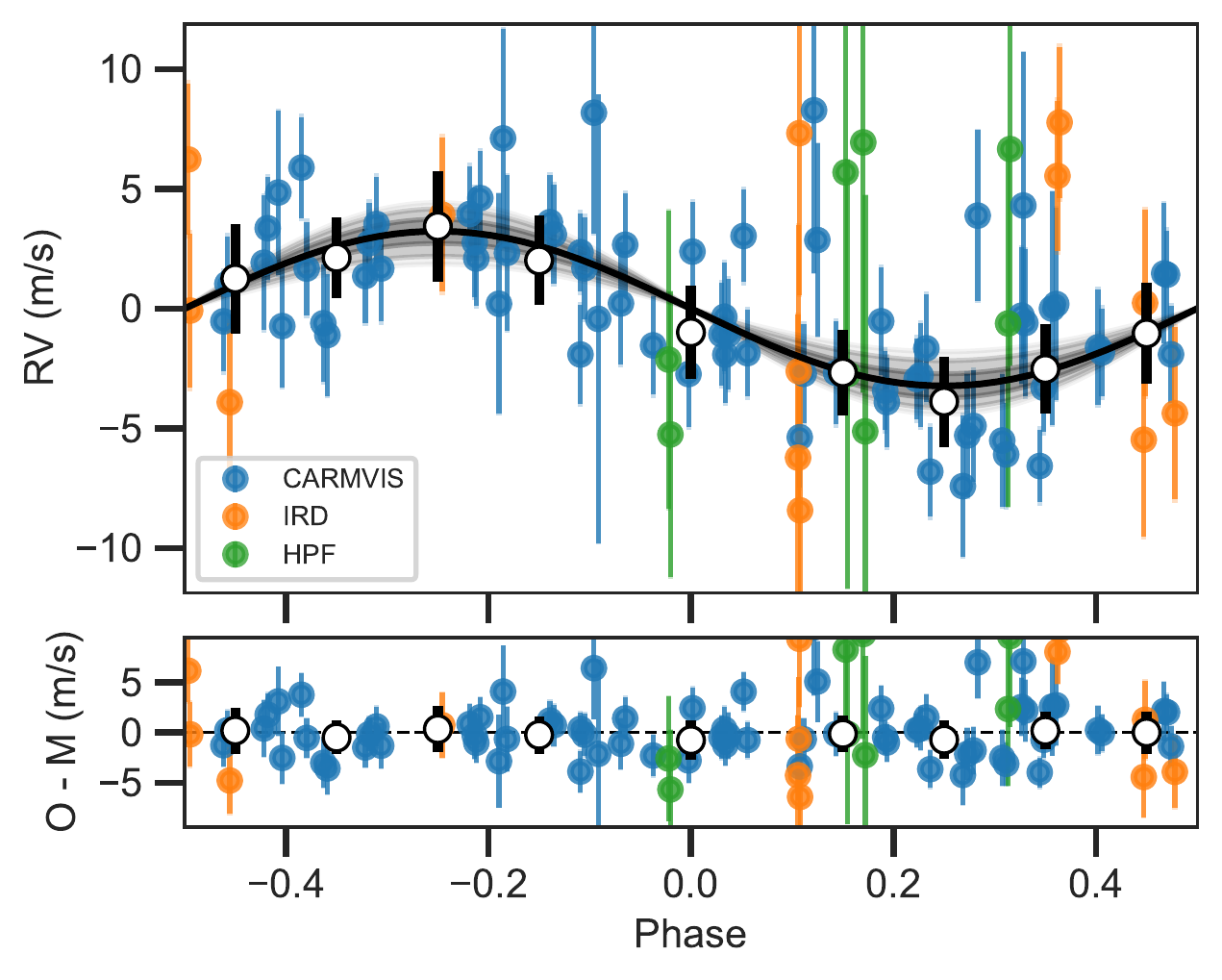}
    \caption{Radial velocities phase folded to the period of the transiting planet after removing the contribution from the quasi-periodic GP. RV data come from CARMENES (blue), IRD (orange), and HPF (green). White circles show data points binned in phase for visualization. The gray shaded area corresponds to the 1, 2, and 3$\sigma$ confidence intervals of the model.
    }
    \label{fig:rvs}
\end{figure}

Finally, to obtain the most precise parameters of the \host{} system, we performed a joint analysis of the \textit{K2}, \textit{TESS}, and ground-based transit photometry, and the RV data, using \texttt{juliet}. The final model is the one discussed in Sect.~\ref{subsec-rv_only}. In addition to modeling the transiting planet with a circular Keplerian orbit, the final model also includes a GP component to model the stellar variability seen in the \textit{K2} light curve and a quasi-periodic GP kernel to model the stellar activity at the second harmonic of the rotational period seen in the RV data. 

Table~\ref{tab:posteriors} shows the parameters fitted in the final joint model, their priors and posterior distributions. The results from the photometry-only part of the fit are fully compatible with the results from \citet{Stefansson2020AJ....159..100S}. Figures~\ref{fig:k2_lc_p1} and \ref{fig:rvs} show the model and residuals of the photometry and RV data, respectively. Table~\ref{tab:derivedparams} lists the transit and physical parameters derived using the stellar parameters in Table~\ref{table-C16_8748-stellar_parameters}.

\section{Discussion}
\label{sec-discussion}

\begin{figure}
    \centering
    \includegraphics[width=\hsize]{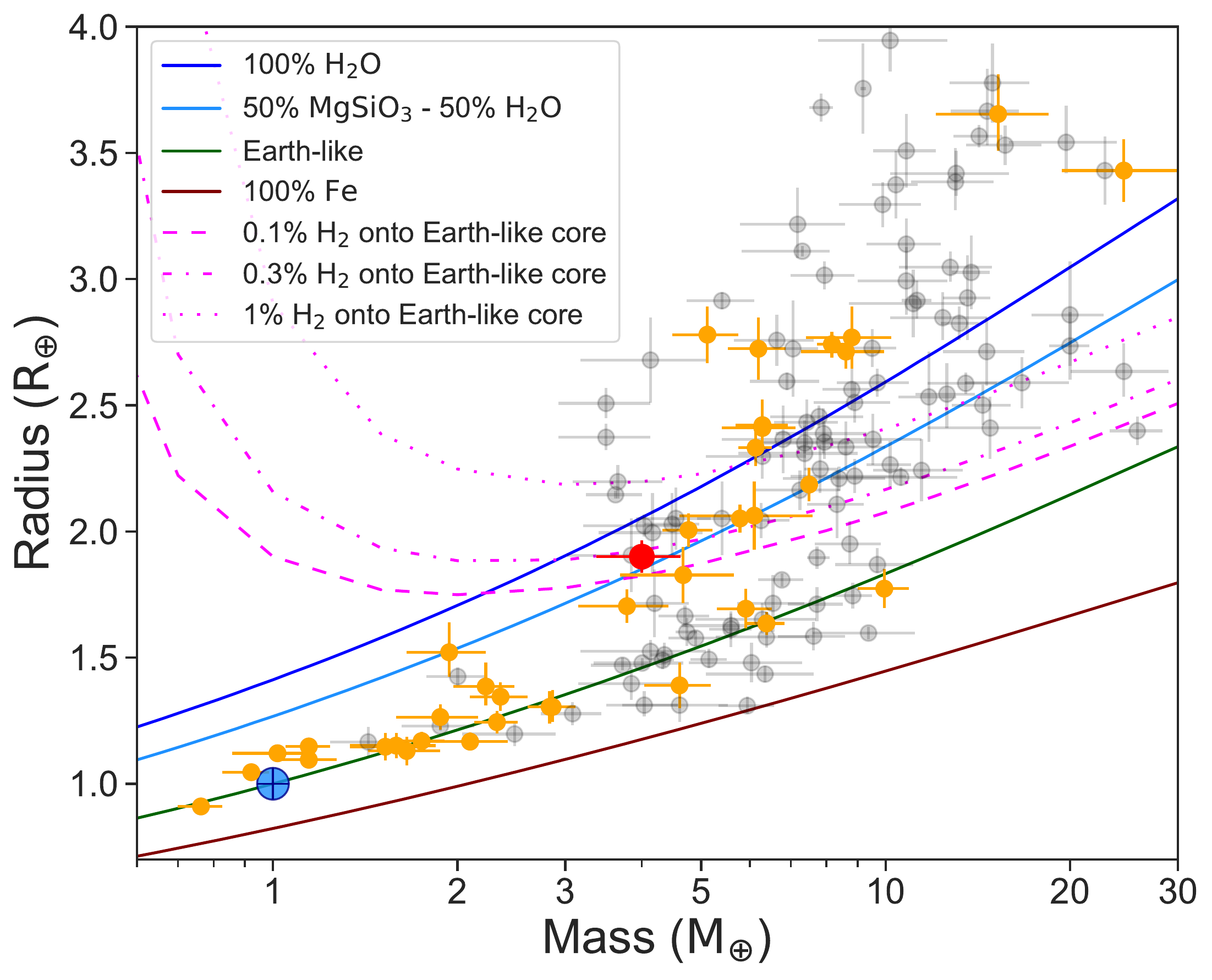}
    \caption{Mass-radius diagram for planets with measurement precision of 25\% (mass) and 8\% (radius). Models from \citet{Zeng2019PNAS..116.9723Z}. Orange: M dwarf hosts, gray: FGK hosts. The red circle indicates the location of G~9-40~b. TEPCat from May 2022.}
    \label{fig:mr}
\end{figure}

\begin{figure}
    \centering
    \includegraphics[width=\hsize]{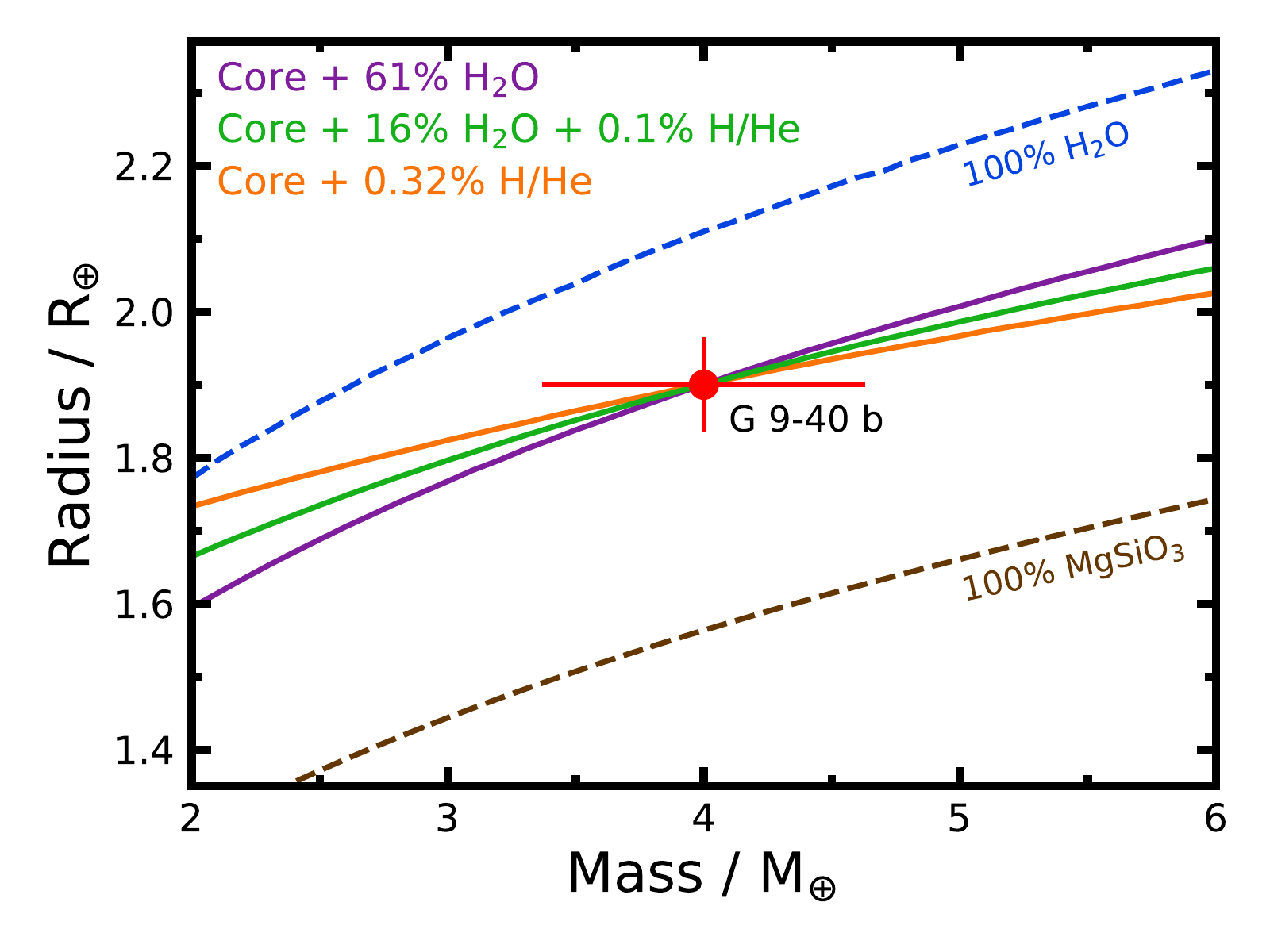}
    \caption{Mass-radius diagram centered on G~9-40~b. The solid purple, orange and green curves show a selection of best-fitting compositions for the planet, assuming an Earth-like structure (1/3 iron, 2/3 silicates), a surface temperature equal to the planetary equilibrium temperature (440.6~K), and a surface pressure of 1~bar. The dashed blue and brown lines represent theoretical pure H$_2$O and pure silicate planets, at the same surface temperature, respectively. Temperature-dependent mass-radius curves were generated using the model of \citet{Nixon2021}.}
    \label{fig:g940b_mr}
\end{figure}

\begin{figure}
    \centering
    \includegraphics[width=\hsize]{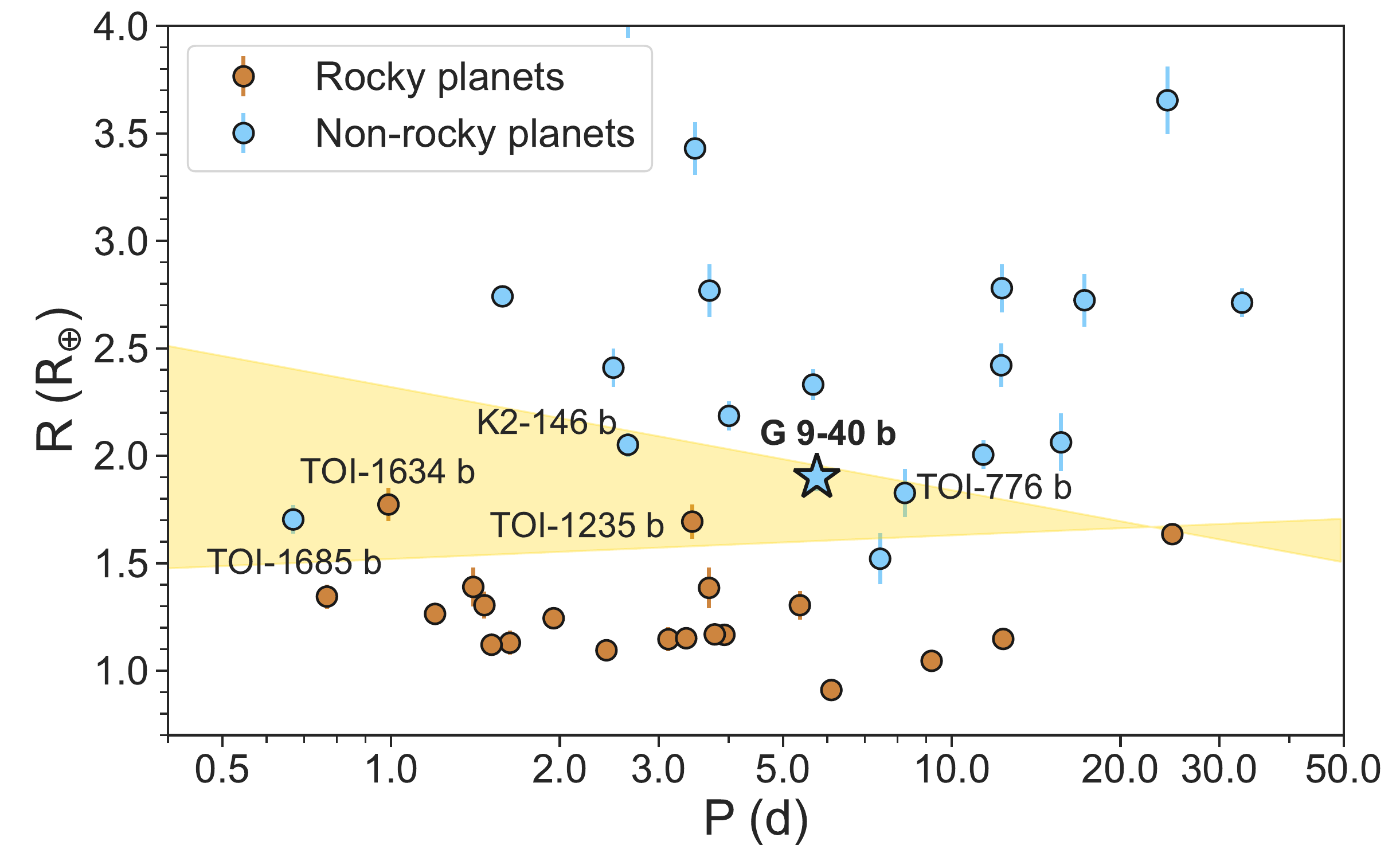}
    \caption{Period-radius diagram for rocky (brown) and non-rocky (blue) planets orbiting M~dwrfs. Planets in the yellow region are referred to as keystone planets \citep{Cloutier2021}. TEPCat from May 2022.}
    \label{fig:radius-gap}
\end{figure}

Thanks to the new \textit{TESS} photometry and CARMENES RV follow-up observations, we are able to characterize the \host{} system and precisely measure the mass of its planet for the first time. \planetb{} is a sub-Neptune planet with a radius of $R_{\rm b} = 1.900\pm0.065\,R_\oplus$ and a mass of $M_{\rm b} = 4.00\pm0.63\,M_\oplus$, resulting in a bulk density of $\rho_{\rm b} = 3.20^{+0.63}_{-0.58}\,\mathrm{g\,cm^{-3}}$. With an equilibrium temperature of $440.6\pm7.6\,\mathrm{K}$, it joins L~98-59~d as the only two warm sub-Neptune planets ($400 < T_{\rm eq} < 600\,\mathrm{K}$) orbiting an M~dwarf with a mass uncertainty at the 15\,\% level or better.

\subsection{Internal composition}
\label{subsec-int-comp}

Figure~\ref{fig:mr} shows the location of \planetb{} in a mass-radius diagram together with the sample of precisely characterized planets --- with a mass uncertainty better than 25\,\% and a radius uncertainty better than 8\,\%, following \citet{Otegi2020} --- orbiting M~dwarfs (orange) and FGK stars (grey). It is evident from Fig.~\ref{fig:mr} that the planet's density is too low for a pure rock composition, and so all viable solutions contain some amount of H$_2$O and/or H/He. Using the internal composition models from \citet{Zeng2019PNAS..116.9723Z} we find that \planetb{} joins a growing population of sub-Neptune planets orbiting M~dwarfs consistent with having a mixture of rock and water ices in 50-50 proportion by mass, also known as water worlds. On the other hand, the mass-radius values of the planet are also consistent with an Earth-like core surrounded by a hydrogen-rich atmosphere of about 0.1--0.3\,\% of its total mass,  assuming a 1\,mbar surface pressure level and an equilibrium temperature of 500\,K. In any case, the size and bulk density of \planetb{} suggests that its internal composition must be different from purely terrestrial.



In order to explore the possible composition of the planet in more detail, we fit internal structure models to the mass, radius, and equilibrium temperature of the planet, considering a wide range of planetary interiors that may be consistent with current observations. The model employs a four-layer structure consisting of a two-component Earth-like core made up of 1/3 iron and 2/3 silicates by mass beneath an envelope consisting of H$_2$O and/or H/He (assuming a solar He fraction, $Y=0.275$). A temperature-dependent equation of state is used for the outer H$_2$O and H/He layers. We assume a nominal surface pressure of 1\,bar and use a temperature profile consisting of an isotherm and an adiabat with the radiative-convective boundary at 10\,bar. For a given mass, composition and set of surface conditions, the model calculates the planet radius. A more detailed description of the model can be found in \citet{Nixon2021}.

We follow a similar approach to the characterization of TOI-776~b and c \citep{Luque2021}, exploring the space of possible values of $x_{\rm core}$, $x_{\rm{H}_2\rm{O}}$ and $x_{\rm{H/He}}$ that are consistent with the bulk properties of G~9-40~b. For each composition we compute radii using a range of masses within 1$\sigma$ of the measured planet mass, in order to find the range of compositions that are consistent with observations.

Given that the planet's bulk density is too low for an entirely rocky composition, we begin by considering end-member scenarios in which the outer layer is made up of either only H$_2$O or only H/He. For the core plus H$_2$O-only models, the mass and radius of G~9-40~b can be explained with a water mass fraction of 44--78\%, with the best-fit solution found at $x_{\rm{H}_2\rm{O}}=0.61$. For the core plus H/He-only scenario, the mass and radius are consistent with a H/He mass fraction of 0.16--0.55\%, with the best-fit solution $x_{\rm H/He}=3.2 \times 10^{-3}$. It is also possible that the planet contains substantial amounts of both H$_2$O and H/He. In this case, a range of $x_{\rm core}$--$x_{\rm{H}_2\rm{O}}$ configurations are consistent with the planet's observed bulk properties, allowing for smaller mass fractions of H/He and H$_2$O than in the pure envelope scenarios. For example, a planet with 0.1\% H/He and 16\% H$_2$O by mass can readily explain the data. However, the overall upper limits remain at $x_{\rm{H}_2\rm{O}} \leq 0.78$ and $x_{\rm H/He} \leq 5.5 \times 10^{-3}$. The best-fit solutions from the internal composition models in each case are shown in Fig.~\ref{fig:g940b_mr}.

According to its orbital period and radius, \planetb{} can be classified as a keystone planet following the definition by \citet{Cloutier2021}. It is located above the radius valley slope measured for low-mass stars ($T_{\rm eff} < 4700\,\mathrm{K}$) from \textit{Kepler/K2} by \citet{CloutierMenou2020} and below the one measured by \citet{VanEylen2021} from a sample of well-studied planets orbiting M dwarfs ($T_{\rm eff} < 4000\,\mathrm{K}$).
Figure~\ref{fig:radius-gap} shows the parameter space occupied by these planets, where \planetb{} joins TOI-1685~b \citep{Hirano2021,Bluhm2021}, TOI-1634~b \citep{Hirano2021,Cloutier2021}, K2-146~b \citep{Hamann2019,Lam2020}, TOI-1235~b \citep{Bluhm2020,Cloutier2020}, and TOI-776~b \citep{Luque2021}. Among them, \planetb{} is the one orbiting the lowest-mass host. Keystone planets lie within the radius valley and are valuable targets to conduct tests of the competing models for the radius valley across a range of stellar masses. To do so, it is necessary to measure their bulk densities and establish their composition. 

Our analysis demonstrates that \planetb{} is inconsistent with being a bare rock and that its composition may range between a water world with a steam atmosphere and a mostly rocky planet with a large hydrogen-rich envelope. Therefore, its location is consistent with the predictions from gas-poor formation models \citep{Lee2014ApJ...797...95L,LeeChiang2016ApJ...817...90L,LeeConnors2021ApJ...908...32L}, but inconsistent with those from atmospheric-mass loss. While \planetb{}, TOI-1685~b, K2-146~b, and TOI-776~b are  all consistent with this scenario, the terrestrial planets TOI-1634~b and TOI-1235~b contradict it. Furthermore, we do not find in our sample the apparent transition between theories as a function of stellar mass suggested by \citet{Luque2021} and \citet{Cloutier2021}. As an example, the ultra-short period planets TOI-1634~b and TOI-1685~b both orbit very similar hosts ($M_\star \sim 0.5\,M_\odot$) despite their differences in internal composition. Enlarging the sample of precisely characterized keystone planets with additional information on the age of the system \citep[e.g.,][]{Berger2020AJ....160..108B,Petigura2022AJ....163..179P,Chen2022AJ....163..249C} and reliable stellar abundances \citep[e.g.,][]{Adibekyan2021Sci...374..330A,DelgadoMena2021A&A...655A..99D} is a way forward towards identifying the dominant mechanism sculpting the radius distribution of small planets orbiting M dwarfs, and to look for differences (if any) with solar-type stars.  



\subsection{Prospects for atmospheric characterization}

\begin{figure}
   \centering
   \includegraphics[width=0.99\hsize]{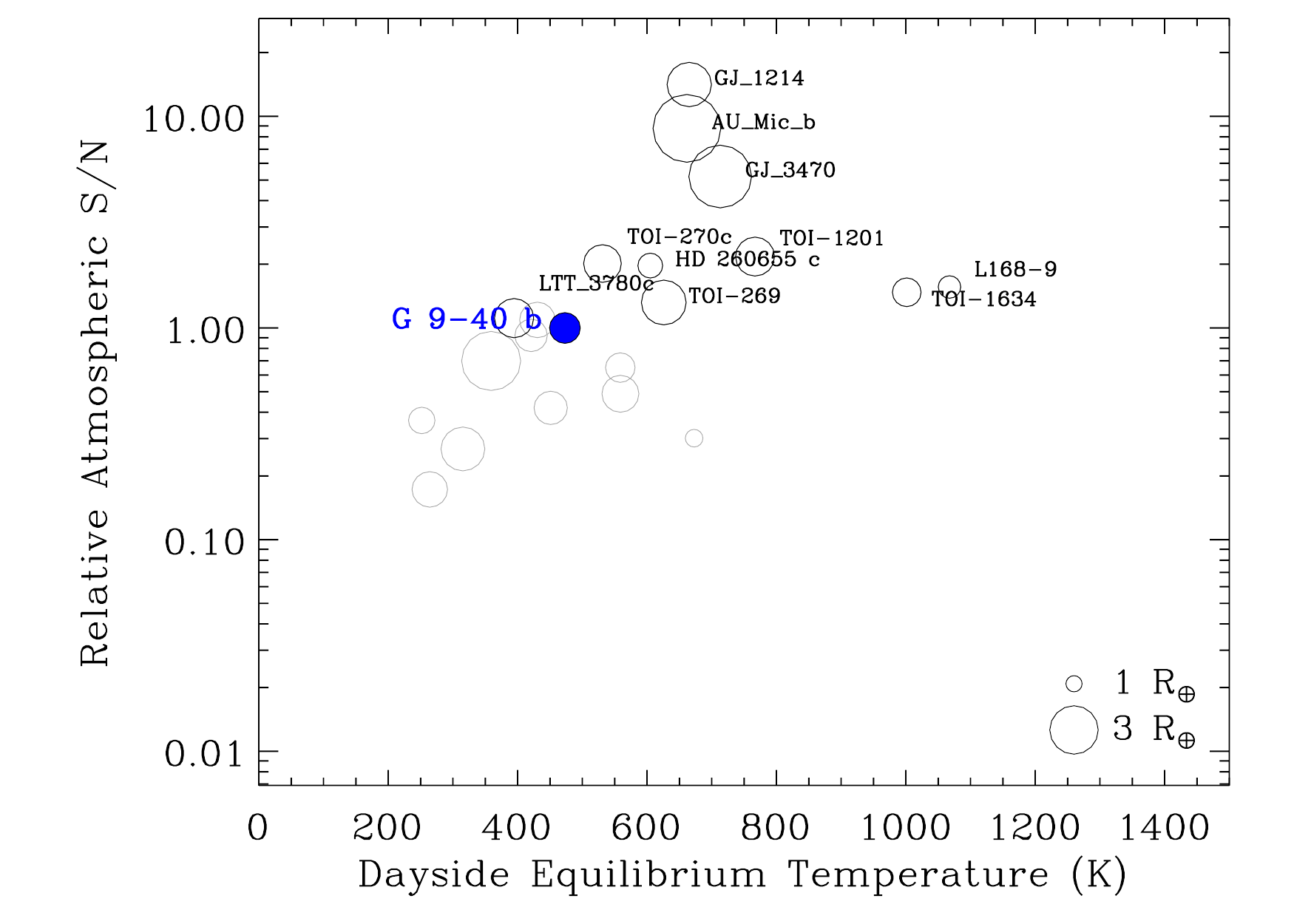}
   \caption{Relative S/N of an atmospheric signal for all sub-Neptune exoplanet candidates (i.e., $3 M_\oplus < M_p < 20 M_\oplus$) orbiting M stars (i.e., $T_{eff} < 3850$ K). The atmospheric characterization S/N is normalized to G 9-40 b, which is highlighted with the filled colored symbol. G 9-40 b is among the most promising exoplanets in this sample for atmospheric characterization, particularly among those with the coolest $T_{eq}$.}
   \label{fig:snrplot}
\end{figure}

\begin{figure*}
   \centering
   \includegraphics[width=0.49\hsize]{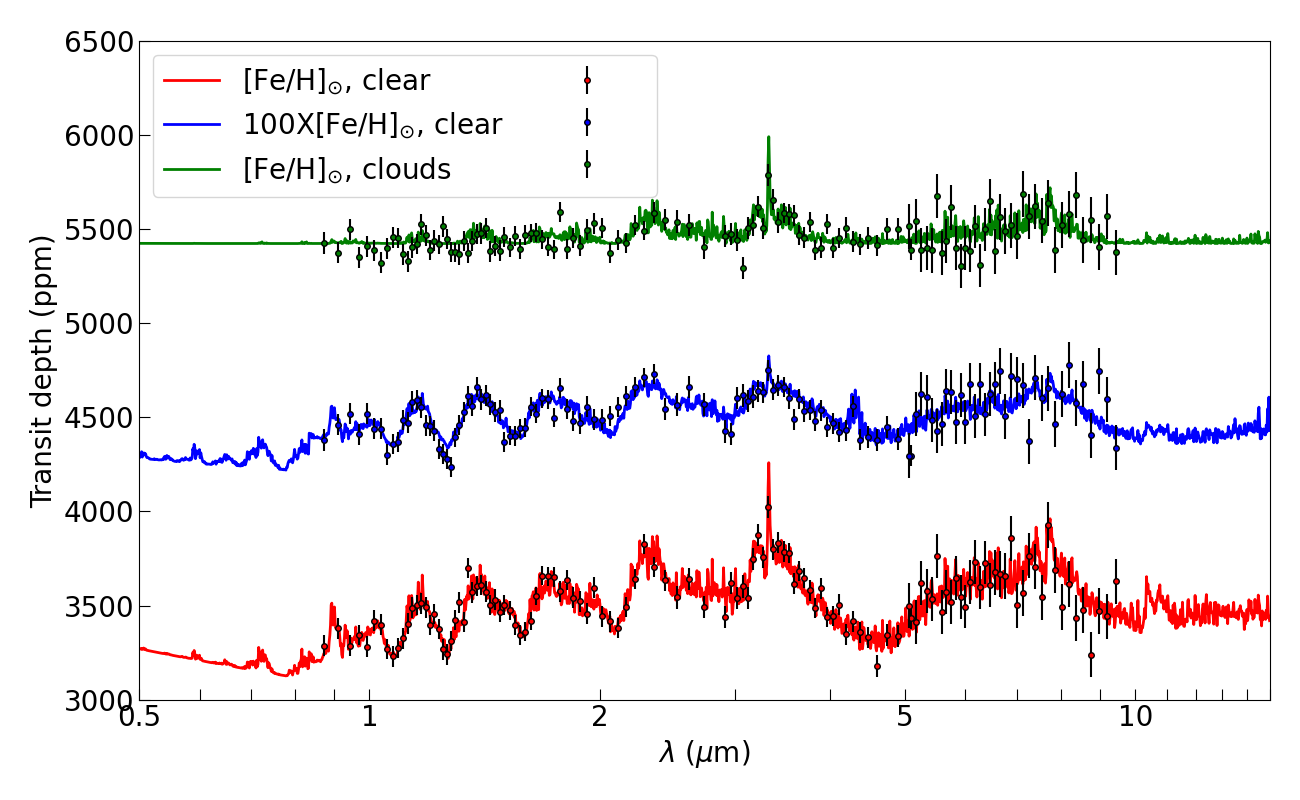}
   \includegraphics[width=0.49\hsize]{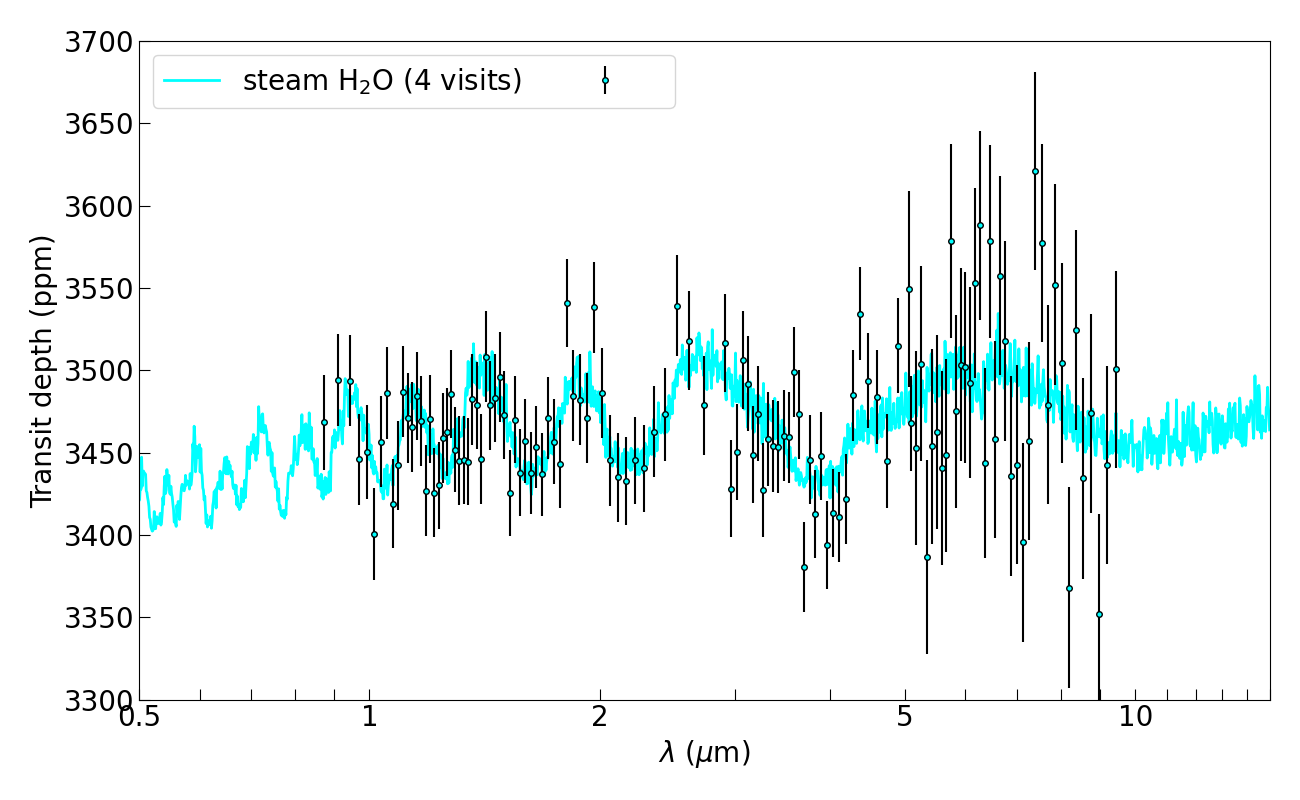}
   \caption{Synthetic transmission spectra of G~9-40 b. Left panel: Fiducial models assuming a cloud-free atmosphere with solar iron abundances (red), cloud-free atmosphere with metallicity enhanced by a factor of 100 (blue, with a vertical offset of +1000 ppm), and atmosphere with solar abundances and optically-thick cloud deck with top pressure of 1\,mbar (green, with a vertical offset of +2000 ppm). Right panel: Model of H$_2$O-dominated atmosphere, without H and He. Estimated uncertainties are shown for the observation of one (left) and four (right) transits with \jwst\ NIRISS-SOSS, NIRSpec-G395M, and MIRI-LRS configurations.}
    \label{Fig:G9_40b_JWSTspectra}
\end{figure*}

Due to the relative brightness of its host star, G~9-40 b is a promising target for upcoming atmospheric studies with the \textit{JWST} \citep{beichman2014}. Figure~\ref{Fig:G9_40b_JWSTspectra} shows the estimated atmospheric S/N (relative to G 9-40 b) of the sample of sub-Neptunes (i.e., $3 M_\oplus < M_p < 20 M_\oplus$) orbiting M stars (i.e., $T_{eff} < 3850$ K). This atmospheric S/N metric is detailed in \citet{2017AJ....154..266N}. It is similar to the transmission spectroscopy metric (TSM) proposed by \cite{kempton2018}, except that it includes the period of the exoplanet and the S/N is calculated for transits over time as opposed to per-transit, like the TSM. Given that many atmospheric characterization observations require multiple transits, the metric shown in Figure~\ref{Fig:G9_40b_JWSTspectra} acknowledges that it is difficult to obtain the required S/N for long-period planets. Nonetheless, both metrics indicate that G~9-40~b is among the few temperate (i.e., $T_{\rm eq} \lesssim 500\,\mathrm{K}$) sub-Neptunes that would be optimal for spectroscopic observations with \jwst.

To further explore the potential of \textit{JWST} transit spectroscopy, we simulated synthetic spectra for a range of atmospheric scenarios, compatible with the internal composition determined in Section \ref{subsec-int-comp}, and instrumental setups. We adopted Tau-REx~III \citep{al-refaie2021} to compute the model atmospheres using the atmospheric chemical equilibrium (ACE) module \citep{agundez2012}, including collisionally induced absorption by H$_2$–H$_2$ and H$_2$–He \citep{abel2011,abel2012,fletcher2018}, and Rayleigh scattering. We show a benchmark model assuming a cloud-free primary atmosphere with solar abundances, and two variants of this model assuming 100$\times$ solar metallicity or an optically thick cloud deck with top pressure of 1\,mbar. Additionally, we modeled a steam atmosphere made of H$_2$O. The physical input parameters have been set to the values reported in Tables \ref{table-C16_8748-stellar_parameters} and \ref{tab:derivedparams}, with the atmosphere assumed to be isothermal at the equilibrium temperature. We note that a higher than solar metallicity may be expected for the atmosphere of G~9-40~b, based on formation models for sub-Neptunes \citep{fortney2013,thorngren2016}. The equilibrium temperature of 440\,K falls within the 300--600\,K range that favors condensation of cloud-forming elements \citep{yu2021}.

The first atmospheric model exhibits molecular absorption features of $\sim$500\,ppm at low spectral resolution. The most prominent features can be attributed to H$_2$O, CH$_4$ and NH$_3$. The model with 100$\times$ solar metallicity has damped absorption features by a factor of $\sim$2, due to a higher mean molecular weight and smaller atmospheric scale-height. The cloudy model also displays smaller absorption features due to the suppression of contributions from deeper atmospheric layers, which is much more effective at shorter wavelengths. A flat spectrum could be observed in case of higher altitude (lower top pressure) clouds. The steam atmosphere presents features of $\sim$100\,ppm.

We used \texttt{ExoTETHyS} \citep{morello2021} to compute binned average spectra, taking into account the spectral response of the \jwst\ instruments, noise scatter and error bars. We simulated \jwst\ spectra for the NIRISS-SOSS (0.6--\SI{2.8}{\micro\metre}), NIRSpec-G395M (2.88--\SI{5.20}{\micro\metre}) and MIRI-LRS (5--\SI{12}{\micro\metre}) instrumental modes. The wavelength bins were specifically determined to have similar counts, leading to nearly uniform error bars per spectral point. In particular, we set a median resolving power of $R\sim 50$ for the NIRISS-SOSS and NIRSpec-G395M modes, and bin sizes of $\sim$0.1--\SI{0.2}{\micro\metre} for the MIRI-LRS. The error bars have been calculated for a single visit of twice the transit duration in each instrumental mode, including the reduction of effective integration time given by the observing efficiency and a factor 1.2 to account for correlated noise. This procedure provides slightly more conservative error bars than those obtained with \texttt{PandExo} \citep{batalha2017}, as already tested in previous studies (e.g., \citealp{espinoza2022}). We obtained error bars of 50--61\,ppm per spectral point for the NIRISS-SOSS and NIRSpec-G395M modes, and 115--122 for the MIRI-LRS bins. These numbers suggest that a single transit observation is sufficient to sample the molecular absorption features in case of a clear atmosphere, even with 100$\times$ solar metallicity. A single NIRSpec-G395M observation could also be sufficient to detect the absorption features in a cloudy scenario, if the top pressure is lower than 1\,mbar (see Fig.~\ref{Fig:G9_40b_JWSTspectra}). The combined information from the NIRISS-SOSS and NIRSpec-G395M modes is crucial to distinguish high-metallicity from cloudy scenarios, owing to the more chromatic damping effect of clouds. While one visit per mode is sufficient to distinguish the selected cases with $>3\sigma$ significance, there can be degeneracies between other configurations. Finally, we estimate that four observations with either NIRISS-SOSS or NIRSpec-G395M modes are necessary to robustly detect the absorption features of a pure H$_2$O steam atmosphere.

We currently lack high S/N observations of sub-Neptune atmospheres, which impede testing our hypotheses about their formation and evolution. Future \jwst\ observations of small sub-Neptunes, such as G~9-40~b, are necessary to increase our understanding on this class of exoplanets.

\section{Summary}
\label{sec-conclusions}

In this work, we determine the dynamical mass of the small planet G~9-40~b using precise RV measurements from CARMENES. Combined with new observations from the \textit{TESS} mission during Sectors 44 to 46, we find the bulk density of the planet, $\rho_{\rm b} = 3.20^{+0.63}_{-0.58}\,\mathrm{g\,cm^{-3}}$, to be inconsistent with a terrestrial composition. From mass and radius measurements alone, our internal structure models are unable to distinguish between a water world with rock and ices mixed in approximately 50-50 proportion by mass and an Earth-like core surrounded by a large H/He envelope contributing about 0.3\% of its total mass. However, future atmospheric observations with \textit{JWST} could break the degeneracies in the internal structure models in just a few visits due to the favorable transmission spectroscopy metrics of the system. The planet joins L~98-59~d as the only two sub-Neptune targets amenable for atmospheric characterization with an equilibrium temperature below 500\,K, at which cloud formation processes are predicted to be less ubiquitous for this type of planets.

G~9-40~b meets the definition of keystone planet coined by \citet{Cloutier2021} according to its location in a period-radius diagram. Among this sample of M-dwarf planets, G~9-40~b is the one orbiting the lowest-mass host, breaking a tentative pattern proposed by \citet{Luque2021} and \citet{Cloutier2021} about the transition between dominant radius valley emergence theories such as atmospheric-mass loss and gas-poor formation as a function of stellar mass. By enlarging the sample of precisely characterized planets in this region of the parameter space combined with further information about the stellar properties, we expect to identify in the near future the dominant mechanism sculpting the radius distribution of small planets orbiting M dwarfs and its differences with solar analogs.

\begin{acknowledgements}
This work was supported by the KESPRINT\footnote{\url{www.kesprint.science}.} collaboration, an international consortium devoted to the characterization and research of exoplanets discovered with space-based missions. 
This paper includes data collected by the Kepler and TESS missions, obtained from the MAST data archive at the Space Telescope Science Institute (STScI). Funding for the Kepler mission is provided by the NASA Science Mission Directorate. Funding for the TESS mission is provided by the NASA Explorer Program. STScI is operated by the Association of Universities for Research in Astronomy, Inc., under NASA contract NAS 5–26555.
We acknowledge the use of public TESS data from pipelines at the TESS Science Office and at the TESS Science Processing Operations Center. Resources supporting this work were provided by the NASA High-End Computing (HEC) Program through the NASA Advanced Supercomputing (NAS) Division at Ames Research Center for the production of the SPOC data products. 
This article is partly based on observations made with the MuSCAT2 instrument, developed by ABC, at Telescopio Carlos S\'{a}nchez operated on the island of Tenerife by the IAC in the Spanish Observatorio del Teide.
CARMENES is an instrument at the Centro Astron\'omico Hispano-Alem\'an (CAHA) at Calar Alto (Almer\'{\i}a, Spain), operated jointly by the Junta de Andaluc\'ia and the Instituto de Astrof\'isica de Andaluc\'ia (CSIC). CARMENES was funded by the Max-Planck-Gesellschaft (MPG), the Consejo Superior de Investigaciones Cient\'{\i}ficas (CSIC), the Ministerio de Econom\'ia y Competitividad (MINECO) and the European Regional Development Fund (ERDF) through projects FICTS-2011-02, ICTS-2017-07-CAHA-4, and CAHA16-CE-3978, and the members of the CARMENES Consortium with additional contributions.
IRD is operated by the Astrobiology Center of the National Institute of Natural Sciences of Japan. 
R.L. acknowledges funding from University of La Laguna through the Margarita Salas Fellowship from the Spanish Ministry of Universities ref. UNI/551/2021-May 26 under the EU Next Generation funds and from the Spanish Ministerio de Ciencia e Innovación, through project PID2019-109522GB-C52, and the Centre of Excellence "Severo Ochoa" award to the Instituto de Astrofísica de Andalucía (SEV-2017-0709).

M.T. is supported by JSPS KAKENHI grant Nos.18H05442, 15H02063, and 22000005.
This work is partly supported by JSPS KAKENHI Grant Numbers JP18H05439, JP21K20376 and JST CREST Grant Number JPMJCR1761.
C.C. acknowledges financial support from the ERDF through project AYA2016-79425-C3-1/2/3-P.
C.M.P. acknowledges the support of the  Swedish National Space Agency (DNR 65/19).
H.J.D. acknowledges support from the Spanish Research Agency of the Ministry of Science and Innovation (AEI-MICINN) under grant PID2019-107061GB-C66.
P.K. acknowledges the support from the grant LTT-20015.
K.W.F.L. was supported by Deutsche Forschungsgemeinschaft grants RA714/14-1 within the DFG Schwerpunkt SPP 1992, Exploring the Diversity of Extrasolar Planets.
This work was supported by the DFG Research Unit FOR2544 ``Blue Planets around Red Stars''
G.M. has received funding from the European Union's Horizon 2020 research and innovation programme under the Marie Sk\l{}odowska-Curie grant agreement No. 895525.

\end{acknowledgements}

\bibliographystyle{aa}
\bibliography{biblio}


\begin{appendix} 


\section{RV time series of best joint fit}

\begin{figure*}
\centering
\includegraphics[width=\hsize]{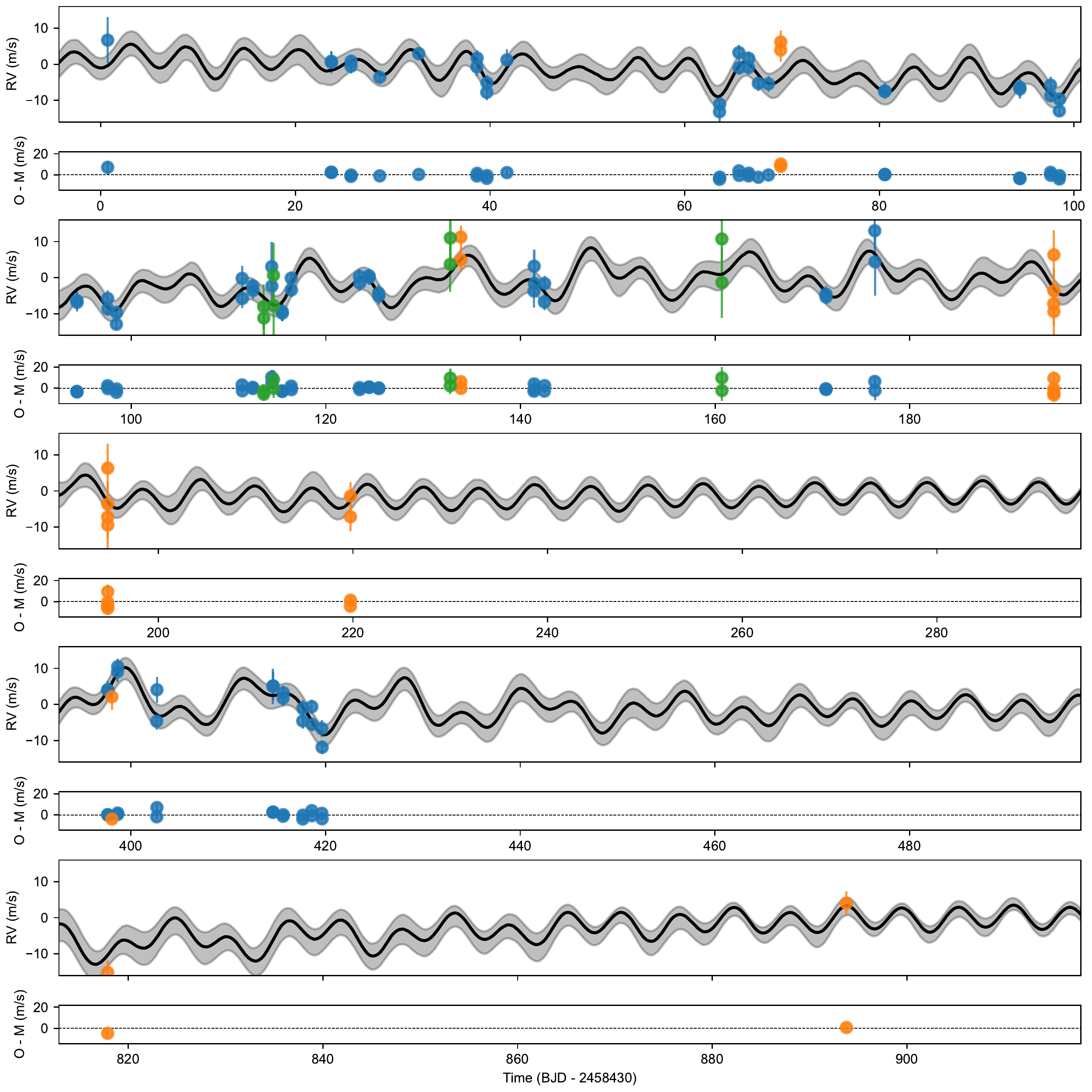}
\caption{RV measurements as a function of time along with the residuals obtained from subtracting our median best joint fit model (black line) and the 68\% posterior band (shown in grey). The color coding of the measurements for each instrument is the same as in Fig.~\ref{fig:rvs}.} 
\label{fig:rv_timeseries1}
\end{figure*}

\end{appendix}

\end{document}